\documentclass[reprint, floatfix,nofootinbib,amsmath,amssymb,unsortedaddress,superscriptaddress]{revtex4-2}
\usepackage[utf8]{inputenc}
\usepackage{graphicx, cancel, booktabs, tabularx, amsmath, amsfonts, amssymb, bm, placeins, float,mathrsfs}
\usepackage{orcidlink}
\usepackage[normalem]{ulem}
\usepackage[table]{xcolor}
\usepackage{multirow}
\usepackage{autobreak}
\usepackage{listings}
\usepackage{xcolor}
\usepackage{hyperref}

\newcolumntype{A}{>{\centering\arraybackslash}m{0.9 cm}}
\newcolumntype{B}{>{\centering\arraybackslash}m{0.5 cm}}
\newcolumntype{D}{>{\arraybackslash}m{6 cm}}

\newcolumntype{E}{>{\centering\arraybackslash}m{1.05in}}
\newcolumntype{L}{>{\centering\arraybackslash}m{0.25in}}
\newcolumntype{M}{>{\centering\arraybackslash}m{0.5in}}
\newcolumntype{N}{>{\centering\arraybackslash}m{1.35in}}

\usepackage{csvsimple}
\usepackage{longtable}
\usepackage{booktabs}

\definecolor{mygreen}{HTML}{1B9E77}
\definecolor{myblue}{HTML}{7570B3}
\definecolor{myred}{HTML}{D73027}
\definecolor{myorange}{HTML}{E6AB02}

\usepackage{lineno}

\begin{document}

\title{Accelerating the Discovery of Materials with Expected Thermal Conductivity via a Synergistic Strategy of DFT and Interpretable Deep Learning}

\author{Yuxuan Zeng
\orcidlink{0009-0004-8954-4377}
}
\affiliation{The Institute of Technological Sciences, Wuhan University, Wuhan 430072, PR China}

\author{Wei Cao
\orcidlink{0000-0001-5131-0728}
}
\email{wei\_cao@whu.edu.cn}
\affiliation{The Institute of Technological Sciences, Wuhan University, Wuhan 430072, PR China}
\affiliation{Key Laboratory of Artiﬁcial Micro- and Nano-Structures of Ministry of Education, School of Physics and Technology, Wuhan University, Wuhan 430072, PR China}

\author{Yijing Zuo}
\affiliation{Key Laboratory of Artiﬁcial Micro- and Nano-Structures of Ministry of Education, School of Physics and Technology, Wuhan University, Wuhan 430072, PR China}

\author{Tan Peng}
\affiliation{Key Laboratory of Artiﬁcial Micro- and Nano-Structures of Ministry of Education, School of Physics and Technology, Wuhan University, Wuhan 430072, PR China}

\author{Yue Hou}
\affiliation{The Institute of Technological Sciences, Wuhan University, Wuhan 430072, PR China}

\author{Ling Miao}
\affiliation{School of Optical and Electronic Information, Huazhong University of Science and Technology, Wuhan 430072, PR China}

\author{Ziyu Wang
\orcidlink{000-0001-9718-1263}
}
\email{zywang@whu.edu.cn}
\affiliation{The Institute of Technological Sciences, Wuhan University, Wuhan 430072, PR China}
\affiliation{Key Laboratory of Artiﬁcial Micro- and Nano-Structures of Ministry of Education, School of Physics and Technology, Wuhan University, Wuhan 430072, PR China}
\affiliation{School of Physics and Microelectronics, Key Laboratory of Materials Physics of Ministry of Education, Zhengzhou University, Zhengzhou 450001, PR China}

\author{Jing Shi}
\affiliation{Key Laboratory of Artiﬁcial Micro- and Nano-Structures of Ministry of Education, School of Physics and Technology, Wuhan University, Wuhan 430072, PR China}

\begin{abstract}
Lattice thermal conductivity (LTC) is a critical parameter for thermal transport properties, playing a pivotal role in advancing thermoelectric materials and thermal management technologies. Traditional computational methods, such as Density Functional Theory (DFT) and Molecular Dynamics (MD), are resource-intensive, limiting their applicability for high-throughput LTC prediction. While AI-driven approaches have made significant strides in material science, the trade-off between accuracy and interpretability remains a major bottleneck. In this study, we introduce an interpretable deep learning framework that enables rapid and accurate LTC prediction, effectively bridging the gap between interpretability and precision. Leveraging this framework, we identify and validate four promising thermal conductors/insulators using DFT and MD. Moreover, by combining sensitivity analysis with DFT calculations, we uncover novel insights into phonon thermal transport mechanisms, providing a deeper understanding of the underlying physics. This work not only accelerates the discovery of thermal materials but also sets a new benchmark for interpretable AI in material science.

\end{abstract}
\date{\today}
\maketitle

\section{Introduction}
\label{sec:Intro}

Lattice thermal conductivity (LTC, $\kappa_{\rm L}$) is a critical physical parameter that quantifies a material's ability to transfer heat through lattice vibrations. It has diverse applications in thermal management~\cite{RN130}, energy conversion~\cite{RN131}, and thermoelectric materials~\cite{RN132}. High-$\kappa_{\rm L}$ materials are used in electronic devices to efficiently dissipate heat and prevent overheating~\cite{RN133}. In contrast, low-$\kappa_{\rm L}$ materials exhibit excellent performance in thermoelectric conversion, making them ideal for developing efficient thermoelectric generators (TEGs)~\cite{RN134} and coolers~\cite{RN135}. $\kappa_{\rm L}$ is crucial for material design and optimization, yet acquiring it for specific materials is challenging.

Traditional experimental methods for measuring $\kappa_{\rm L}$, such as the laser flash method~\cite{RN137} and thermal conductivity probes~\cite{RN138}, are often inefficient. In theoretical calculations, solving the Boltzmann Transport Equation~(BTE) based on Density Functional Theory (DFT) is regarded as the most reliable method for determining $\kappa_{\rm L}$~\cite{RN140,RN139}. Molecular Dynamics~(MD) offers an alternative~\cite{RN141}; however, the former is limited by its substantial computational resource requirements, while the latter's accuracy depends on the choice of interatomic potentials~\cite{RN142}. In recent years, machine learning (ML) has emerged as a powerful and efficient data mining tool, gaining widespread application in materials science~\cite{RN144}. The intersection of materials science and artificial intelligence is commonly referred to as ``Materials Informatics''~\cite{RN143} or the ``Materials Genome''~\cite{RN145}. Early ML-based predictions of material properties primarily aimed at achieving high prediction accuracy~\cite{RN146,RN285,RN147}. Nevertheless, such efforts, relying on ``black-box'' models, offered limited support for advancing theoretical research in materials science. The focus has shifted towards model interpretability, prompting greater adoption of ``white-box'' models that contribute more substantially to theoretical advancements~\cite{RN148}. Research in materials informatics related to thermal conductivity has followed this trend. Efforts to predict $\kappa_{\rm L}$ using black-box models such as Random Forest~(RF)~\cite{RN152}, Gaussian Process Regression (GPR)~\cite{RN151}, and eXtreme Gradient Boosting (XGBoost)~\cite{RN150} have yielded reliable accuracy, whereas, the interpretability of these models is often hindered by their complexity. Alternatively, Genetic Programming-based Symbolic Regression (GPSR)~\cite{RN153} algorithms offer better interpretability, but their simplicity comes at the cost of reduced accuracy. The prevailing view is that the behavior and processes of complex models are difficult to understand and interpret, while simple models often lack strong fitting capabilities~\cite{RN155,RN154,RN148}. This presents a challenging trade-off, but the situation is gradually improving. The Sure Independence Screening and Sparsifying Operator (SISSO)~\cite{RN157}, based on compressed sensing and symbolic regression, has not only surpassed the accuracy of the Slack semi-empirical model~\cite{RN158} but also narrowed the gap with black-box models like Kernel Ridge Regression (KRR) and GPR, and has been used to quantify feature sensitivity and identify key physical parameters influencing $\kappa_{\rm L}$~\cite{RN156}. Despite its potential, SISSO's applicability is constrained by its high computational demands. Enumerating combinations of features and operators to construct descriptors poses an NP-hard problem~\cite{RN157}, with resource requirements increasing for high-dimensional input features~\cite{RN159}.
\begin{figure*}
    \centering
    \includegraphics[width=0.8\textwidth]{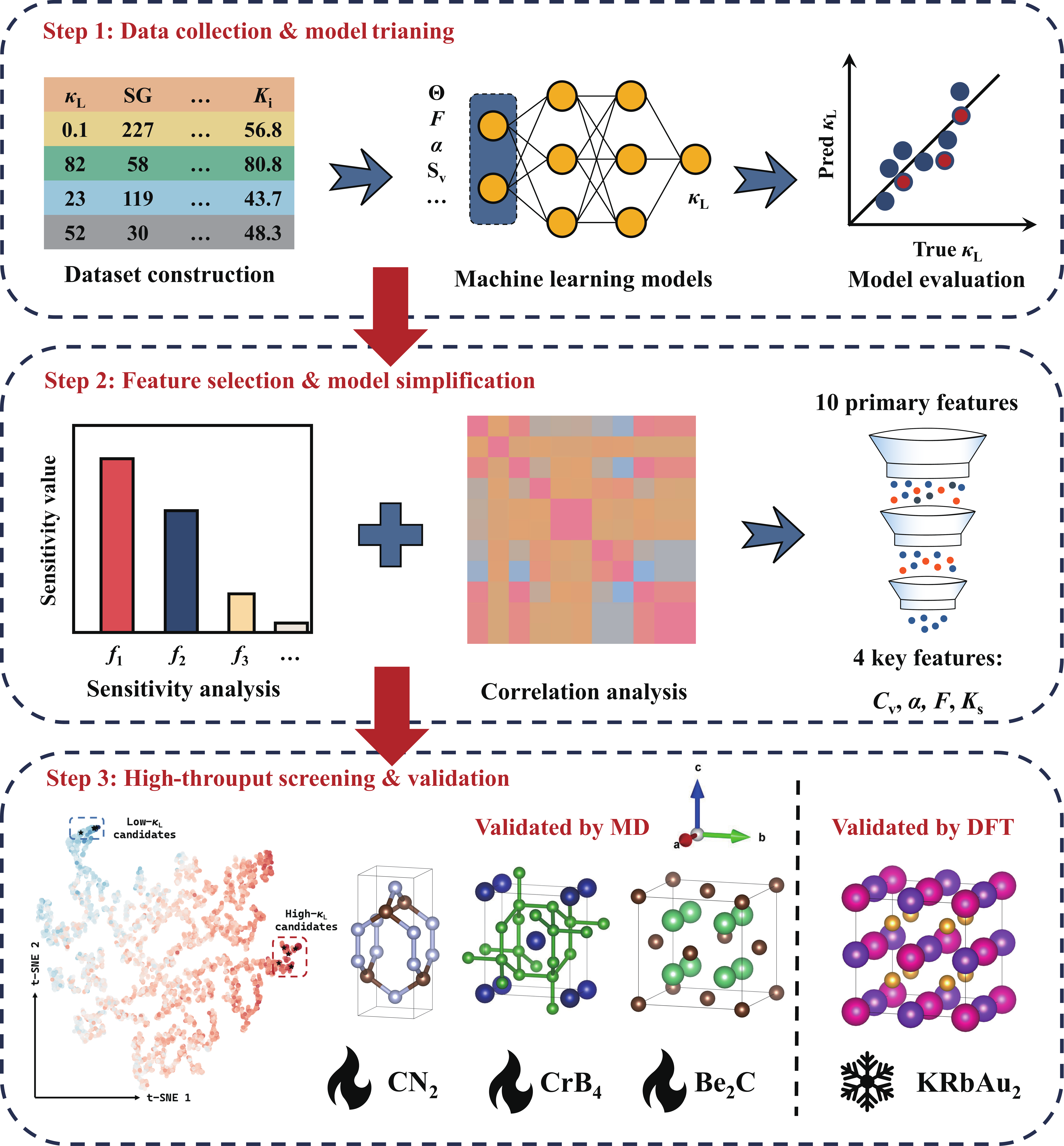}
    \caption{A schematic illustration of the DL framework for $\kappa_{\rm L}$ modeling and high-throughput prediction. This framework facilitates accurate and interpretable predictions of $\kappa_{\rm L}$ for inorganic crystalline materials.}
    \label{fig:ga}
\end{figure*}

Kolmogorov-Arnold Networks (KANs)~\cite{RN160} , a novel neural network architecture distinct from Multi-Layer Perceptrons (MLPs), show significant promise in solving ordinary differential equations~\cite{odes}, time series classification~\cite{dong2024kolmogorov}, and fluid modeling~\cite{PENG2024100184}. In some cases, smaller KAN models outperform MLPs~\cite{RN165,RN160}. 
Notably, their capabilities in symbolic regression offer interpretability advantages that absent in MLPs~\cite{RN165,RN160}. As previously mentioned, existing work has successfully used black-box models to accurately predict $\kappa_{\rm L}$. However, achieving accurate $\kappa_{\rm L}$ predictions while simultaneously visualizing the model's decision-making process, akin to white-box model, remains a significant challenge. The emergence of KANs provides an opportunity to address this issue.

Fig.~\ref{fig:ga} illustrates the framework of our work, which aims to model and predict $\kappa_{\rm L}$ using the white-box deep learning model KAN and compare its performance with that of conventional black-box and white-box ML models. In addition to accuracy, we perform sensitivity analysis to examine interpretability differences across various white-box models and to assess the contributions of individual features to $\kappa_{\rm L}$. This approach further elucidates the relationships between $\kappa_{\rm L}$ and key features, as well as feature-feature correlations. 
Additionally, Crystal Graph Convolutional Neural Network (CGCNN)~\cite{RN321} is employed to predict critical physical features required for $\kappa_{\rm L}$ modeling with high accuracy, forming a two-stage framework that significantly reduces the time required for screening potential materials. This framework enables rapid qualitative assessments of thermal insulators and conductors. 
Furthermore, validated through DFT and MD calculations, we successfully identified three thermal conductors and one thermal insulator using this framework. Finally, taking two of these materials as case studies, we leverage DFT to provide in-depth insights into their respective phonon thermal transport mechanisms. 

\section{Methods}

\subsection{Kolmogorov-Arnold Networks (KANs)}

The Kolmogorov-Arnold representation theorem~\cite{RN337} provides that any multivariate continuous function defined on a bounded domain can be expressed as a finite composition of continuous functions of a single variable, combined with the operation of addition~\cite{RN167}. For a differentiable function $f:{{\left[ 0,1 \right]}^{n}}\to \mathbb{R}$:
\begin{align}
    f\left( \mathbf{x} \right)=f\left( {{x}_{1}},...,{{x}_{n}} \right)=\sum\limits_{q=1}^{2n+1}{{{\Phi }_{q}}\left( \sum\limits_{p=1}^{n}{{{\phi }_{q,p}}\left( {{x}_{p}} \right)} \right)}
    \label{eq:8}
\end{align}

\noindent
where ${{\phi }_{q,p}}:\left[ 0,1 \right]\to \mathbb{R}$ as well as ${{\Phi }_{q}}:\mathbb{R}\to \mathbb{R}$. In this context, $p$ denotes the number of top operators, $q$ denotes the number of bottom operators, and $n$ denotes the number of nodes in the bottom network (which corresponds to the dimensionality of the input features in the input layer), $\mathbf{x}=\left( {{x}_{1}},...,{{x}_{n}} \right)$ represents feature vector.

In the original K-A representation theorem, the number of nonlinear layers is limited to 2, and the number of hidden layer nodes is set to $2n+1$, which fixes the network structure to $\left[ n,2n+1,1 \right]$. However, Liu et al.~\cite{RN160} are not constrained by these limitations. In KANs, both the number of layers and the width of the network are arbitrary, which enhances the feasibility of the K-A representation theorem for ML applications. Therefore, for a KAN with ${{n}_{0}}$-D inputs, L layers, and ${{n}_{\text{L}}}=1$-D output, a more precise definition is:
\begin{widetext}
\begin{align}
    \text{KAN}\left( \mathbf{x} \right) =\sum\limits_{{{i}_{L-1}}=1}^{{{n}_{L-1}}}{{{\phi }_{L-1,{{i}_{L}},{{i}_{L-1}}}}\left( \sum\limits_{{{i}_{L-2}}=1}^{{{n}_{L-2}}}{\cdots \left( \sum\limits_{{{i}_{2}}=1}^{{{n}_{2}}}{{{\phi }_{2,{{i}_{3}},{{i}_{2}}}}\left( \sum\limits_{{{i}_{1}}=1}^{{{n}_{1}}}{{{\phi }_{1,{{i}_{2}},{{i}_{1}}}}\sum\limits_{{{i}_{0}}=1}^{{{n}_{0}}}{{{\phi }_{0,{{i}_{1}},{{i}_{0}}}}\left( {{x}_{{{i}_{0}}}} \right)}} \right)} \right)\cdots } \right)}
    \label{eq:9}
\end{align}
\end{widetext}
the expression can be simplified to:
\begin{align}
    \text{KAN}\left( \mathbf{x} \right)=\left( {{\mathbf{\Phi }}_{L-1}}\circ {{\mathbf{\Phi }}_{L-2}}\circ \cdots \circ {{\mathbf{\Phi }}_{1}}\circ {{\mathbf{\Phi }}_{0}} \right)\mathbf{x}
    \label{eq:10}
\end{align}

\noindent
in the expression, ${{\mathbf{\Phi }}_{i}}=\left\{ {{\phi }_{q,p}} \right\}$ denotes the i-th layer of the KAN, defined as a tensor of 1-D activation functions ${{\phi }_{q,p}}$. In the implementation details, $\phi \left( x \right)$ is represented as a linear combination of the basis function $b\left( x \right)$ and B-spline function:
\begin{align}
    \phi \left( x \right) &=\omega \left[ b\left( x \right)+\text{spline}\left( x \right) \right] \notag \\
    &=\omega \left[ \frac{x}{1+{{\text{e}}^{-x}}}+\sum\limits_{i}{{{c}_{i}}{{B}_{i}}\left( x \right)} \right]
    \label{eq:11}
\end{align}

\noindent
where B-spline function is defined as:
\begin{align}
    B_{i,0}(x) &=
    \begin{cases} 
    1, & \text{if } t_i \leq x < t_{i+1}, \\
    0, & \text{otherwise.}
    \end{cases}
    \notag
    \\ B_{i,k}(x) &= 
    \frac{x - t_i}{t_{i+k} - t_i} B_{i,k-1}(x) + \frac{t_{i+k+1} - x}{t_{i+k+1} - t_{i+1}} B_{i+1,k-1}(x). 
    \label{eq:eq12}
\end{align}

In KANs, the coefficients $c$, the degree $k$, and the grid $G$ of the spline functions are all learnable parameters, which are updated through backpropagation.

\subsection{Sure Independence Screening \& Sparsifying Operator (SISSO)}

SISSO constructs a descriptor vector ${{\mathbf{d}}_{n}}$ using features and operators, and models the target vector $\mathbf{P}$ through a linear combination of $n$-dimensional descriptors. The vector of primary features, ${{\mathbf{\Phi }}_{0}}$, serves as the starting point for constructing descriptors. The operator set is defined as:
\begin{align}
    {{\hat{H}}^{\left( \text{m} \right)}}\equiv \left\{ I,+,-,\times ,\div ,\exp ,\log ,\text{abs},\sqrt{{}}{{,}^{-1}}{{,}^{-2}}{{,}^{-3}} \right\}\left[ {{\phi }_{1}},{{\phi }_{2}} \right],
    \label{eq:eq13}
\end{align}
where ${{\phi }_{1}}$ as well as ${{\phi }_{2}}$ are terms in ${{\mathbf{\Phi }}_{i}}$. The superscript $^{\left( \text{m} \right)}$ indicates that SISSO retains only descriptors with physical meaning. For example, features that are added or subtracted must have the same dimensions, and features involved in logarithmic or square root operations cannot be negative. The new features constructed during the $i$-th iteration can be expressed as
\begin{align}
    {{\Phi }_{i}}=\bigcup\limits_{k}{{{{\hat{h}}}_{k}}^{\left( \text{m} \right)}\left[ {{{\hat{\phi }}}_{i}},{{{\hat{\phi }}}_{j}} \right]},\forall {{\hat{h}}_{k}}^{\left( \text{m} \right)}\in {{\hat{H}}^{\left( \text{m} \right)}}\text{ and }\forall {{\phi }_{i}},{{\phi }_{j}}\in {{\Phi }_{i-1}},
    \label{eq:eq14}
\end{align}
where ${{\hat{h}}_{k}}^{\left( \text{m} \right)}$ represents single operator of ${{\hat{H}}^{\left( \text{m} \right)}}$, ${{\phi }_{i}}$ and ${{\phi }_{j}}$ are different elements from ${{\Phi }_{i-1}}$. SIS~\cite{RN263} constructs the feature subspace with the highest correlation to the target $\mathbf{P}$ through vector inner products, while SO selects the top $n$ features with the highest relevance using regularization techniques to form the descriptor matrix ${{\mathbf{d}}_{n}}$. The linear coefficients can be approximately determined by solving the equation ${{\mathbf{c}}_{n}}={{\left( \mathbf{d}_{n}^{\text{T}}{{\mathbf{d}}_{n}} \right)}^{-1}}\mathbf{d}_{n}^{\text{T}}\mathbf{P}$, so the model constructed by SISSO can be represented as:
\begin{align}
    \mathbf{\hat{P}}={{\mathbf{d}}_{n}}{{\mathbf{c}}_{n}}={{c}_{0}}+{{c}_{1}}{{d}_{1}}+\cdots +{{c}_{n}}{{d}_{n}},
    \label{eq:eq15}
\end{align}
where $\mathbf{\hat{P}}$ represents the estimated target vector of the model.

\subsection{Sobol indices}

Sobol is a method for feature importance analysis based on analysis of variance (ANOVA)~\cite{RN270}. It allocates a portion of the total variance to each input variable or its interactions with other variables, thereby providing valuable information about the importance of each input variable~\cite{RN264}. The first-order and total Sobol indices can be defined as follows:
\begin{align}
    {{S}_{i}}=\frac{\text{Var}\left[ {{\text{E}}_{\sim i}}\left[ Y|{{X}_{i}} \right] \right]}{\text{Var}\left[ Y \right]},
    \label{eq:eq16}
\end{align}
\begin{align}
    S_{i}^{\text{T}}=\frac{{{\text{E}}_{\sim i}}\left[ \text{Var}\left[ Y|{{\mathbf{X}}_{\sim i}} \right] \right]}{\text{Var}\left[ Y \right]}.
    \label{eq:eq17}
\end{align}

For a specific value of ${{X}_{i}}$, the value of $Y$ can be determined by averaging the model evaluations over a sample of ${{\mathbf{X}}_{\sim i}}$ while keeping ${{X}_{i}}=x_{i}^{*}$ fixed, where ${{\mathbf{X}}_{\sim i}}$ represents all variables except ${{X}_{i}}$.

\subsection{Kucherenko indices}

Unlike metrics such as Sobol, LIME, and SHAP, the Kucherenko indices accounts for the dependencies among input features. As an extension of the Sobol sensitivity indices, the Kucherenko indices is specifically designed to quantify the sensitivity of model outputs to input variables while considering these interdependencies~\cite{RN314,RN264,RN274}. The Kucherenko indices essentially involves computing Sobol indices after first employing Copulas~\cite{RN276} (typically Gaussian Copulas) to separate each feature's marginal distribution from the dependency structure~\cite{RN275}, as detailed in Supplementary Table~1. This approach approximates the correlated variables as independent before calculating the Sobol indices. The Gaussian copula is constructed based on the standard normal distribution, and its mathematical expression is:
\begin{align}
    C\left( {{u_1},{u_2}, \ldots ,{u_d};\Sigma } \right) = {\Phi _\Sigma }\left[ {{\Phi ^{ - 1}}\left( {{u_1}} \right),{\Phi ^{ - 1}}\left( {{u_2}} \right) \ldots ,{\Phi ^{ - 1}}\left( {{u_d}} \right)} \right],
    \label{eq:eq18}
\end{align}
where $ u_i = F_i(x_i) $ is the Cumulative Distribution Function (CDF) of the marginal distribution, $ \Phi^{-1} $ is the inverse of the CDF of the standard normal distribution $ \mathcal{N}(0, 1) $, $ \Phi_{\Sigma} $ is the CDF of a multivariate normal distribution with mean zero and covariance matrix $ \Sigma $, and $ \Sigma $ is the correlation matrix, describing the dependence structure between the random variables.

In Kucherenko indices, the first-order index $K_1$ quantifies the extent to which the variance of the target depends on the variance of a feature in isolation, while the total-effect index $K^{\rm T}$ captures the influence of that feature under interaction effects. A simple example is $y = x_1 + \cos(x_2) + \sin(x_1x_2)$, where the sensitivity of the terms $\{x_1, \cos(x_2)\}$ with respect to $y$ is reflected by $K_1$, whereas the interaction term $\{\sin(x_1x_2)\}$ is captured by $K^{\rm T}$. Notably, ${K_1 > K^{\rm T}}$ indicates a strong correlation with other features, while $K^{\rm T} = 0$ signifies an exact dependence on other inputs~\cite{RN156,RN264}.

\subsection{Dataset construction \& feature preprocessing}

The dataset used in this work is sourced from the aflowlib.org database~\cite{RN277}, comprising a total of 5,578 entries. For feature selection, the focus was primarily on characteristics related to vibrational, thermodynamic, and mechanical properties at room temperature (300K), which are theoretically associated with $\kappa_{\rm L}$ according to first-principles researches~\cite{RN279,RN278}. For more details, please refer to Table~\ref{tab:2}.
In the modeling aimed at $\kappa_{\rm L}$, the model predicts the logarithm of $\kappa$ ($\log(\kappa_{\rm L})$) rather than the $\kappa_{\rm L}$ itself. This is because the logarithmic transformation compresses the target value space, thereby enhancing the performance of the ML models~\cite{RN280}.
For deep models such as MLPs and KANs, normalization of features is essential as these models rely on gradient descent for parameter optimization. Normalization enhances numerical stability by ensuring that features with different ranges contribute equally to the gradient computations; otherwise, features with larger values might dominate the optimization process~\cite{RN281}. However, this step is unnecessary for XGBoost~\cite{RN282}, as it is based on decision trees. Additionally, for interpretability reasons, feature normalization is often omitted in recent works based on SISSO~\cite{RN157,RN156,RN283}. In this work, we employed Min-Max Normalization, which is expressed as:
\begin{align}
    x' = \frac{{x - {x_{\min }}}}{{{x_{\max }} - {x_{\min }}}},
    \label{eq:eq19}
\end{align}
where $x$ is the primary feature value, ${{x}_{\min }}$ and ${{x}_{\max }}$ are the minimum and maximum values of the feature, respectively, and ${x}'$ is the normalized feature value.

\subsection{Implementation of DFT \& GPUMD methods}

DFT and AIMD calculations are performed using VASP~\cite{RN350,RN349}. The Projector-Augmented Wave (PAW) method~\cite{RN351} and the Perdew-Burke-Ernzerhof (PBE) functional~\cite{RN352} within the Generalized Gradient Approximation (GGA) are employed for electron exchange-correlation. The cutoff energy is set to 500 eV, and the electronic convergence threshold is $10^{-8}$ eV. For the calculation of $\kappa_{\rm L}$, both the phonon BTE method and MD simulations are used. For the stable structures, $\kappa_{\rm L}$ are computed using phono3py~\cite{togo2015distributions,togo2023implementation,chaput2013direct}. 
While $\kappa_{\rm L}$ of unstable structures are determined through Graphics Processing Units Molecular Dynamics (GPUMD)~\cite{RN356} AIMD-NVT simulations are carried out in the canonical ensemble (NVT) with a Nos\'{e}--Hoover thermostat at 300~K for a duration of 5~ns. A uniformly spaced sampling strategy was adopted, from which 250 frames were selected for training and another 250 frames for testing the NeuroEvolution Potential (NEP) model.

The NEP model employs a feed-forward neural network to establish a mapping between local descriptors and the atomic site energies~\cite{wu2024correcting}. In a single hidden layer neural network comprising $N_{\rm neu}$ neurons, the energy $U_i$ of atom $i$ is expressed as:
\begin{align}
    U_i = \sum_{\mu = 1}^{N_{\rm neu}} 
    \boldsymbol{\omega}_{\mu}^{(1)} 
    \tanh \left( 
        \sum_{\nu=1}^{N_{\rm des}} 
        \boldsymbol{\omega}_{\mu v}^{(0)} q_{\nu}^i - \mathbf{b}_{\mu}^{(0)} 
    \right) 
    - b^{(1)}.
\end{align}

Here, $N_{\rm neu}$ denotes the number of descriptor components, and $q_{\nu}^i$ represents the $\nu$-th descriptor component of the $i$-th atom. ${\boldsymbol{\omega}}_{\mu\nu}^{\left(0\right)}$ and ${\boldsymbol{\omega}}_\mu^{\left(1\right)}$ are the weight matrix from the input layer to the hidden layer and the weight vector from the hidden layer to the output node, respectively. $\mathbf{b}_{\mu}^{\left(0\right)}$ and $b^{\left(1\right)}$ correspond to the bias vector in the hidden layer and the bias applied to the node $U_i$, respectively. $\tanh{\left(x\right)}$ denotes the nonlinear activation function in the hidden layer~\cite{fan2022gpumd}. The training of the NEP potential is carried out by optimizing the free parameters through the minimization of a loss function, defined as the weighted sum of the RMSEs of the energies, forces, and virial stresses. The non-equilibrium molecular dynamics (HNEMD) method is employed in GPUMD simulations. In the HNEMD method, an additional driving force ($F_{\rm e}$) is applied to each atom:
\begin{align}
    \mathbf{F}_i^{\rm ext}=F_{\rm e}\cdot\mathbf{W}_i,
\end{align}
where $F_{\rm e}$ is a driving force parameter with the dimension of inverse length, and $\mathbf{W}_i$ is the virial tensor, defined as:
\begin{align}
    \mathbf{W}_i=\sum_{j \neq i}\mathbf{r}_{ij}\otimes\frac{\partial U_j}{\partial \mathbf{r}_ji}.
\end{align}

$U_j$ is the energy of atom $j$ at the given site. The driving force parameter should be systematically tested 3 and a $F_{\rm e}$ value of $1\times 10^{-4} {\AA}^{-1}$ has been tested and found to be optimal for this study. After the driving force was applied to generate the homogeneous heat current, the system was further equilibrated in the NVT ensemble until thermal equilibrium was achieved. The cumulative time average of the instantaneous thermal conductivity, as defined by Eq.~(\ref{eq:kappa_avg}), was then calculated to obtain $\kappa_{\rm avg}(t)$:
\begin{align}
    \kappa_{\rm avg}(t)=\frac{1}{t}\int_{0}^{t}\kappa\left(\tau\right)d\tau.
    \label{eq:kappa_avg}
\end{align}

Specific heat and phonon dispersions are calculated using Phonopy~\cite{RN355}, while the pCOHP analysis is carried out with the LOBSTER code~\cite{nelson2020lobster}. Structural visualizations and ELF analyses are performed using VESTA~\cite{momma2011vesta}. Furthermore, to complement these results, we present additional validation cases based on Machine Learning Interatomic Potentials (MLIPs)~\cite{ne-mlip} in Supplementary Note~7.

\section{Results}

\subsection{Establishment and evaluation of $\kappa_{\rm L}$ models}
\label{model-training}

In selecting black-box models, we consider XGBoost and MLP as our primary options. XGBoost, a representative of ensemble learning, has demonstrated notable performance in various applications, including electrocaloric temperature change prediction in ceramics~\cite{RN285} and materials mechanical property prediction~\cite{RN286}. XGBoost is an enhancement of gradient boosting decision trees (GBDT). Compared to traditional GBDT, XGBoost introduces several innovations, including regularization (to improve generalization), a two-step gradient approximation for the objective function (to enhance computational efficiency), column subsampling (to reduce noise and boost generalization), and handling missing values (using a default direction for tree nodes, making it suitable for sparse datasets)~\cite{RN261}. MLP is a type of feedforward neural network consisting of an input layer, one or more hidden layers, and an output layer. Each neuron in a given layer is fully connected to all neurons in the preceding layer through weight matrices, allowing the network to perform linear combinations and nonlinear mappings of high-dimensional features. Neurons in each hidden layer typically use nonlinear activation functions such as ReLU, which endow the MLP with considerable expressive power, enabling it to approximate arbitrarily complex nonlinear functions. The training process of an MLP is carried out using the backpropagation algorithm, which optimizes the network's weights through gradient descent to minimize a loss function. While MLPs exhibit substantial performance advantages in handling complex data patterns and feature learning tasks, their computational complexity and sensitivity to hyperparameter choices can make the training process time-consuming~\cite{RN287}.
\begin{figure}[h]
    \centering
    \includegraphics[width=\columnwidth]{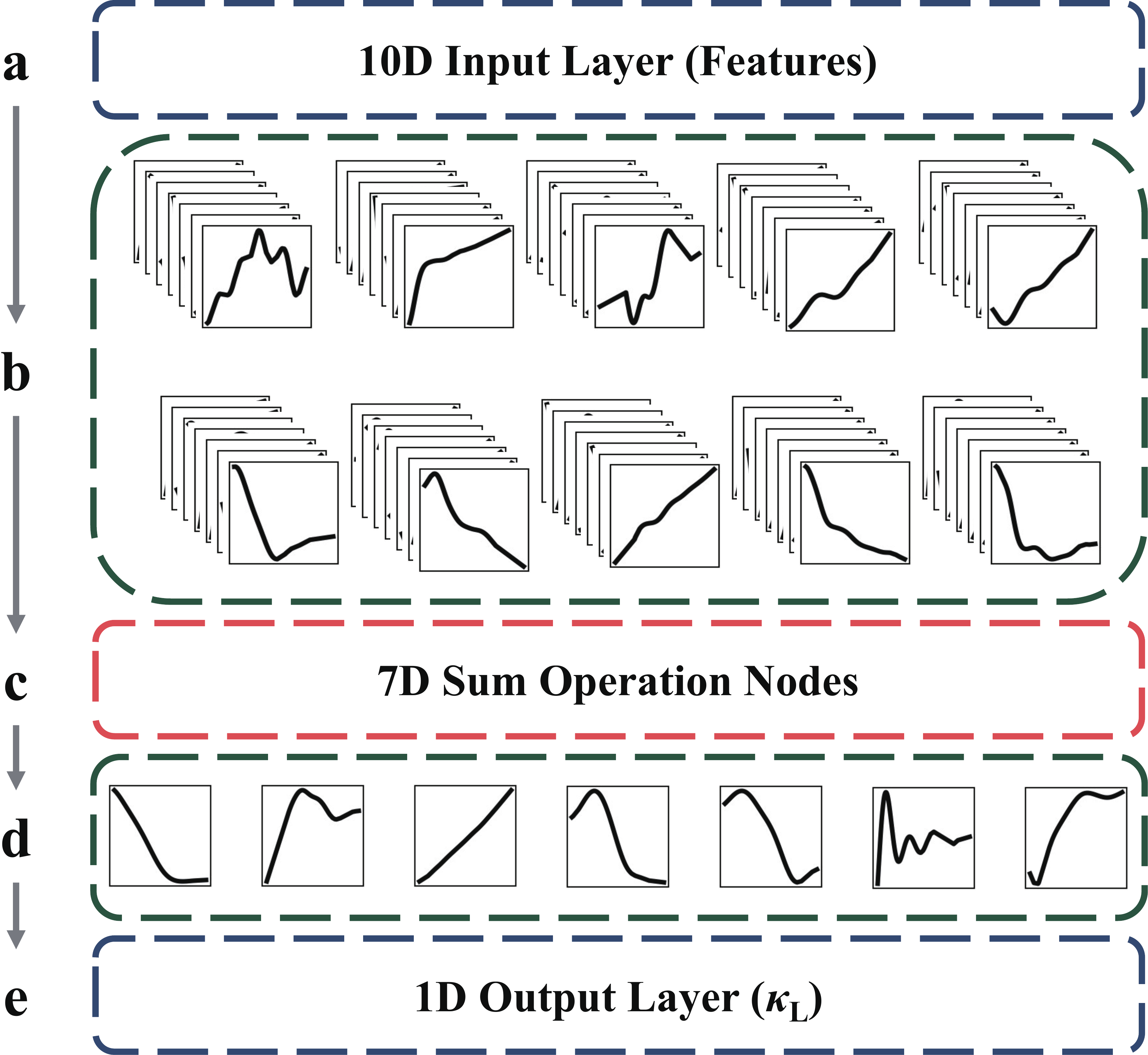}
    \caption{Architecture of the KAN model based on the full dataset. (a) Input layer; (b) First KAN layer with a total of 70 residual activation functions; (c) Sum-operation nodes layer that connects the two KAN layers and performs summation over the outputs of the first KAN layer; (d) Second KAN layer with 7 residual activation functions; (e) Output scalar corresponding to the predicted value of $\log(\kappa_{\rm L})$.}
    \label{fig:arc}
\end{figure}
\begin{table}[h]
    \caption{The ranges for hyperparameter optimization.}
    \centering
    \begin{tabular}{lll}
        \toprule
        \textbf{Model}       & \textbf{Hyperparameter} & \textbf{Domain/Value}     \\ 
        \midrule
        \multirow{4}{*}{MLP} & hidden\_sizes           & {[}64, 64, 64{]} (fixed) \\
                             & learning\_rate          & {[}0.001, 0.1{]}         \\
                             & batch\_size             & {[}64, 256{]}            \\
                             & num\_epochs             & {[}100, 600{]}           \\ 
        \midrule
        \multirow{7}{*}{XGB} & n\_estimators           & {[}50, 500{]}            \\
                             & learning\_rate          & {[}0.0001, 0.5{]}        \\
                             & max\_depth              & {[}1, 6{]}               \\
                             & subsample               & {[}0.8, 1{]}             \\
                             & colsample\_bytree       & {[}0.8, 1{]}             \\
                             & reg\_alpha              & {[}0.8, 1{]}             \\
                             & reg\_lambda             & {[}5, 50{]}              \\ 
        \midrule
        \multirow{6}{*}{KAN} & width                   & {[}5, 15{]}              \\
                             & grid                    & {[}5, 15{]}              \\
                             & k                       & {[}2, 10{]}              \\
                             & lamb\_l1                & {[}5, 50{]}              \\
                             & steps                   & {[}20, 35{]}             \\
                             & lr                      & {[}0.01, 1.0{]}          \\ 
        \botrule
    \end{tabular}
    \label{tab:1}
\end{table}

In our case, KAN was implemented using the pykan-0.0.5 library (\url{github.com/KindXiaoming/pykan}), while MLP and XGBoost were built using the PyTorch~\cite{RN288} and scikit-learn~\cite{RN289} libraries, respectively. Initially, a trial-and-error approach was employed to determine approximate ranges for the hyperparameters that might yield good performance for KAN, MLP, and XGBoost. Subsequently, we employed the Optuna~\cite{RN290} library to perform automated hyperparameter optimization for these models. The task of identifying optimal hyperparameters for ML models can be conceptualized as finding the optimal solution to a multivariate optimization problem. It is crucial to acknowledge that hyperparameter optimization often converges to local optima rather than the global optimum, with achieving a global optimum remaining an inherently challenging endeavor. Within the hyperparameter space, multiple local optima may exist, and no algorithm can guarantee that the solution obtained is globally optimal~\cite{RN291}. Nevertheless, it is generally observed that with sufficient iterations, optimization algorithms often produce models with comparable performance across different hyperparameter configurations. This suggests that while achieving global optimality remains a complex challenge, practical implementations typically deliver satisfactory results within reasonable computational effort. The hyperparameter optimization ranges are detailed in Table~\ref{tab:1}, while the final selected values are presented in Supplementary Table~2.

\begin{figure}[h]
    \centering
    \includegraphics[width=\columnwidth]{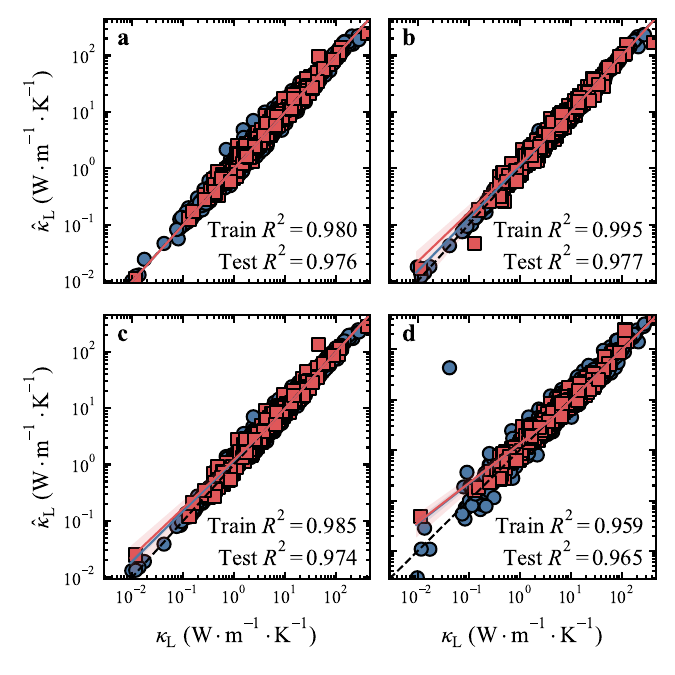}
    \caption{Prediction performance of various models upon the full dataset (Table~\ref{tab:2}). (a) KAN; (b) XGB; (c) MLP; and (d) SISSO. Blue circular markers indicate training samples, while red square markers represent test samples. In each parity plot, the $x$-axis denotes the reference values of $\kappa_{\rm L}$, and the $y$-axis shows the corresponding model predictions.}
    \label{fig:performance}
\end{figure}

In contrast to the previously discussed algorithms, SISSO imposes the highest computational demands during model training. Therefore, we followed hyperparameter configurations recommended in established methodologies from related work~\cite{RN156}. To enhance interpretability, we maintained strict dimensional consistency across all features in SISSO, effectively eliminating any potential for invalid operations.

For regression models, the coefficient of determination ($R^2$) provides more information compared to other commonly used criteria~\cite{RN296}. As shown in Fig.~\ref{fig:performance}, on the full dataset, XGB slightly outperforms KAN, which in turn outperforms MLP; however, the differences are marginal, and all three models are obviously better than SISSO. Additionally, SISSO exhibits severe errors in a very limited subset of samples with extremely low $\kappa_{\rm L}$ values (approximately $\log(\kappa_{\rm L})<-1$). This phenomenon indicates that SISSO struggles with datasets characterized by class imbalance. SISSO tends to fit more to samples with a higher frequency and a narrower range of target values when dealing with imbalanced data, leading to suboptimal extrapolation performance. As a form of symbolic regression, SISSO inherently faces challenges in addressing sample imbalance. Common approaches to mitigate this issue in symbolic regression include resampling~\cite{RN298} and weighting~\cite{RN298,RN297}. However, these methods inevitably impact the model's interpretability~\cite{RN299,RN300}. Therefore, we chose not to compromise SISSO's interpretability solely for the purpose of enhancing its performance. It is noteworthy that XGBoost outperforms the deep learning models in this experiment, which can be attributed to several factors: 
\begin{itemize}
    \item Chemical features are often non-smooth, whereas deep models tend to favor smooth solutions~\cite{RN303,RN304};
    \item Different dimensions of molecular features typically carry distinct information, yet deep models tend to integrate features across dimensions~\cite{RN304};
    \item Tree-based models like XGBoost inherently handle redundant features, while deep models are more susceptible to interference from such features~\cite{RN303};
    \item MLP is rotation-invariant, and any rotation-invariant learning process inherently exhibits the worst-case sample complexity~\cite{RN303}.
\end{itemize}

In addition to accuracy, the symbolic models constructed by SISSO and KAN are also worth considering. The symbolic model provided by SISSO is as follows:
\begin{align}
    \log{(\kappa_{\rm L})_{{\rm{SISSO}}}} = & - 2.8 + 0.3\log \frac{{{\Theta ^2}}}{{\left| {{C_{\rm{p}}} - {C_{\rm{v}}}} \right|}} \notag \\ 
     &- 0.17 \times {10^7}\gamma {K_{\rm{i}}}\sqrt[3]{{\Theta {S_{\rm{v}}}}}.
    \label{eq:1}
\end{align}

The linear coefficients in Eq.~(\ref{eq:1}) are obtained through least-square regression. For the specific physical meanings of the features, please refer to Table~\ref{tab:2}. Our dataset consists of a total of 10 features, but the SISSO model utilizes only seven of these. Some features, such as the space group number, which are typically not considered to be related to $\kappa_{\rm L}$. SISSO effectively identifies and excludes such extraneous variables. However, it overlooks physical quantities that are theoretically linked to $\kappa_{\rm L}$, such as the thermal expansion coefficient~\cite{RN302,RN301}, while feature sensitivity analysis indicates that the thermal expansion coefficient is regarded as an important parameter for modeling $\kappa_{\rm L}$ (see Section~\ref{interpretability}).
\begin{table*}
    \caption{Features adopted in the full dataset.}
    \centering
    \begin{tabular}{llll}
        \toprule
        \textbf{Definition}                              & \textbf{Notation} & \textbf{Unit}                              & \textbf{Domain} \\ 
        \midrule
        The space group number for the relaxed structure  
                                                     & ${\rm SG}$                  & -                                   & - \\  
        Debye temperature                             & $\Theta$              & K                                          & [24.9909, 2144.93] \\  
        Gr{\"u}neisen parameter                           & $\gamma$              & -                                   & [0.00210114, 3.86399] \\  
        Heat capacity per cell at constant volume (300K)  
                                                     & ${{C}_{\text{v}}}$    & $k_{\rm B}$                          & [1.11566, 275.775] \\  
        Heat capacity per cell at constant pressure (300K)  
                                                     & ${{C}_{\text{p}}}$    & $k_{\rm B}$                          & [0.0, 286.301] \\  
        Thermal expansion coefficient (300K)         & $\alpha$              & K$^{-1}$                                   & [$-3.17 \times 10^{-6}$, $5.66601 \times 10^{-4}$] \\  
        Vibrational entropy per atom (300K)          & ${{S}_{\text{v}}}$    & meV/K                         & [0.0165489, 1.03614] \\  
        Vibrational free energy per atom (300K)      & $F$                   & meV                                   & [$-233.268$, 210.145] \\  
        Static bulk modulus (300K)                   & ${{K}_{\rm s}}$    & GPa                                        & [0.0, 440.158] \\  
        Isothermal bulk modulus (300K)               & ${{K}_{\text{i}}}$    & GPa                                        & [0.0, 437.329] \\  
        \botrule
    \end{tabular}
    \label{tab:2}
\end{table*}

In this experiment, the network structure of the KAN model was also determined through automatic hyperparameter optimization via Optuna. The final architecture selected a dimensionality of 7 for the sum operation nodes (i.e., the network structure is  $10 \times 7 \times 1$), with one KAN layer preceding and another following this operation. In the first KAN layer, which connects to the input layer, there are $10 \times 7 = 70$  residual activation functions, while in the second KAN layer, connected to the output layer, the number is $7 \times 1 = 7$. The detailed network architecture is shown in Fig.~\ref{fig:arc}. The MLP we designed features hidden layers with dimensions of $64 \times 64 \times 64$. This architecture balances robust pattern recognition capabilities and mitigating issues such as gradient explosion or vanishing gradients. Despite achieving comparable performance, KAN and MLP exhibit significant differences in terms of parameter count, with MLP generally requiring a substantially higher number of parameters. 
The analytical expressions provided by KAN are not equivalent to the model obtained through data training, unlike SISSO, where the resulting symbolic formula is exactly the model itself. In the KAN Layers, each residual activation function consists of a basis function and a linear combination of several B-spline functions. To symbolize the model, KAN selects the elementary functions with the highest linear correlation to the nodes as substitutes. After symbolization, KAN undergoes further training to determine the affine parameters for each symbolic function. Although this process enhances interpretability, it may slightly reduce performance. Nevertheless, in this experiment, the $R^2$ value of the symbolic formula derived from KAN, at 0.9655, remains higher than that obtained by SISSO. The symbolic model constructed by KAN is detailed in Supplementary Information Eq.~(\ref{eq:1}). The analytical expression derived by KAN is much more complex than that of SISSO, as each feature is involved in 7 operations, a consequence of the KAN architecture. Please note that each feature in this formula has undergone Min-Max normalization as part of the feature preprocessing. However, this preprocessing step does not affect the model's interpretability, as it is easy to revert~\cite{RN305}. Despite the fact that the analytical expressions provided by KAN and SISSO accurately capture the relationship between features and $\kappa_{\rm L}$, there are a few points to consider:
\begin{itemize}
    \item The symbolic models constructed through machine learning require more physical features (e.g., SISSO involves 7 features, while KAN utilizes all 10 features) compared to semi-empirical models (such as Slack~\cite{RN158}, which, when the temperature variable is fixed at 300K, requires only 5 additional features). Moreover, obtaining the corresponding features for unknown materials often involves high experimental or computational costs, making such models appear less appealing at present.
    \item In materials inverse design, it is often necessary to regulate dominant features to induce changes in the target physical properties. Although the decision-making process of symbolic models is transparent, relying solely on the symbolic model itself to infer the dominance (or ``contribution'') of different features is impractical.
    \item Both KAN and SISSO achieve reliable predictive accuracy, but they differ in terms of interpretability. For instance, due to dimensional constraints (see Supplementary Table~2), SISSO excludes certain features during model construction, and we cannot guarantee that the excluded features are necessarily redundant or useless. Similarly, in KAN, while different features have varying weights, we cannot ensure that all features genuinely influence $\kappa_{\rm L}$. Therefore, it is essential to introduce effective and robust quantitative methods to objectively evaluate the interpretability of these models.
\end{itemize}

\subsection{Deepening physical interpretability through sensitivity analysis}
\label{interpretability}
We identified the most critical features for $\kappa_{\rm L}$ through two steps. The first step is correlation analysis, achieved by calculating correlation coefficients. The second step is sensitivity analysis, accomplished by computing sensitivity indices.

\begin{figure*}
    \centering
    \includegraphics[width=\textwidth]{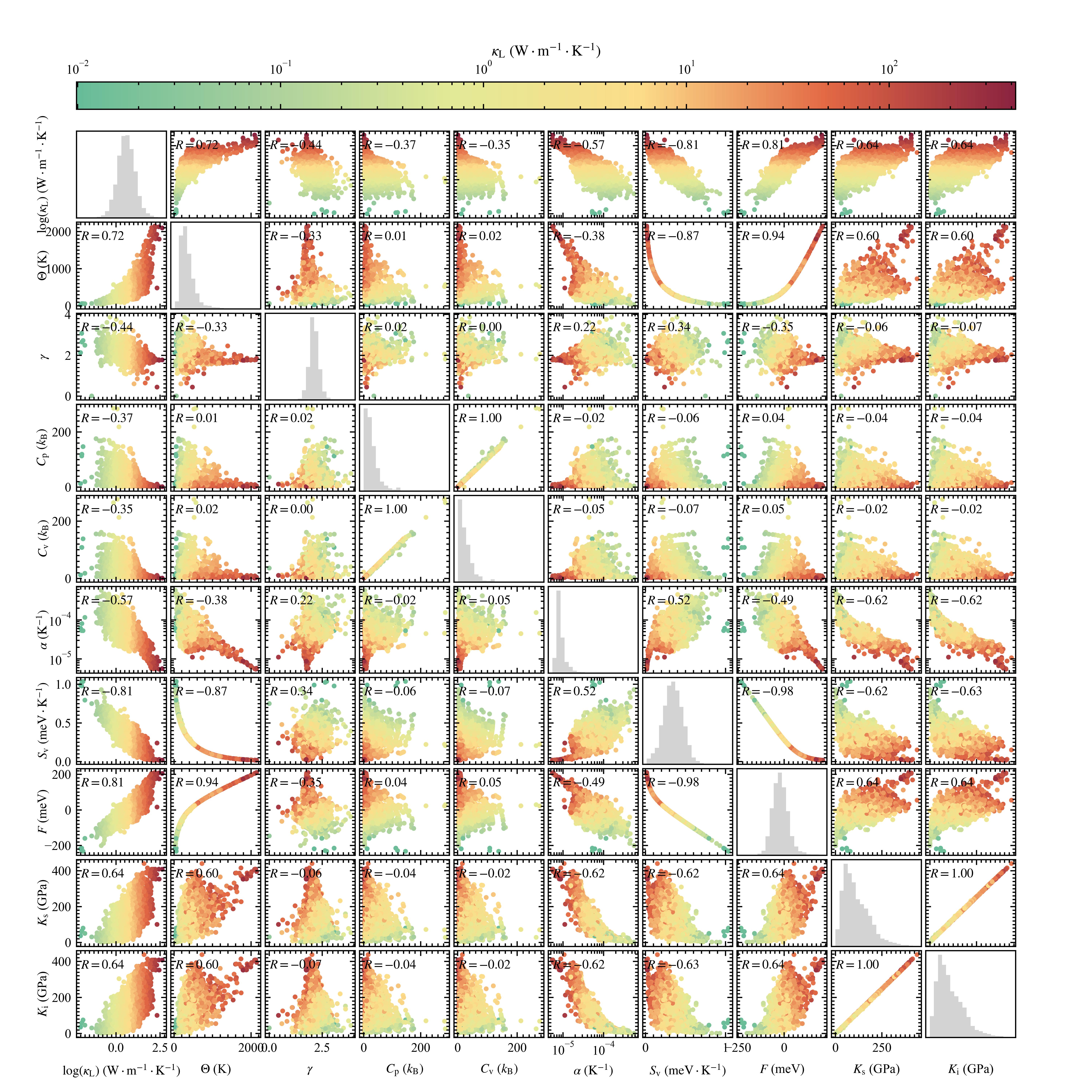}
    \caption{Correlation matrix scatterplot displaying pairwise relationships among key thermal and mechanical properties of materials. Diagonal panels show histograms of individual variable distributions. Off-diagonal scatterplots are color-coded by $\kappa_{\rm L}$ on a logarithmic scale, highlighting how $\kappa_{\rm L}$ varies with respect to each feature pair. Pearson correlation coefficients ($R$) quantify linear relationships between variables, aiding interpretation of interdependencies relevant for thermal transport analysis.}
    \label{fig:cor-map}
\end{figure*}

\begin{figure}[h]
    \centering
    \includegraphics[width=\columnwidth]{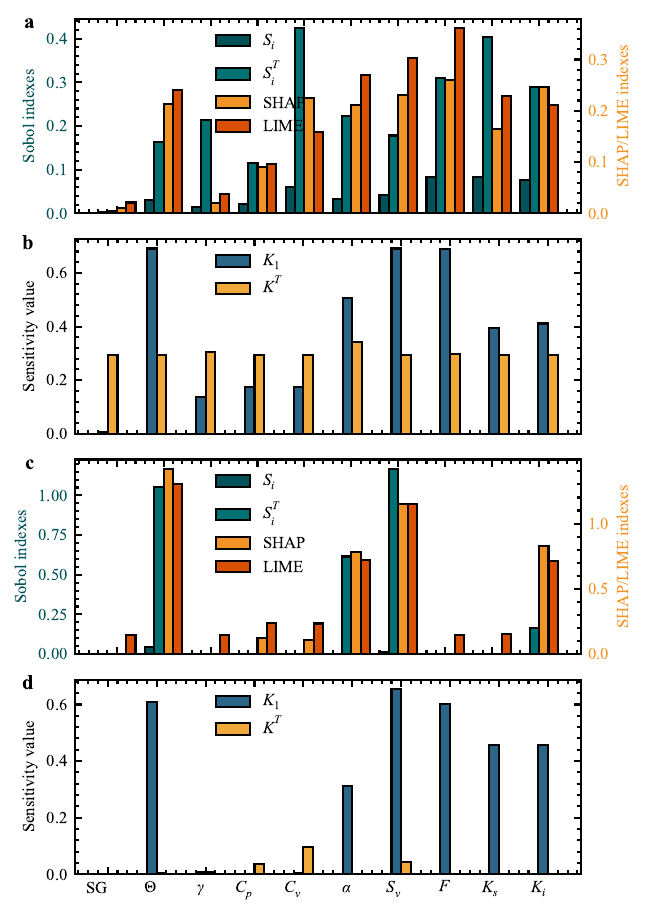}
    \caption{Feature sensitivities of the KAN and SISSO surrogate models. (a) and (b) are based on the KAN model, while (c) and (d) correspond to the SISSO model. Specifically, (a) and (c) show results obtained using SHAP, LIME, and Sobol indices, whereas (b) and (d) display sensitivities calculated using the Kucherenko index. The $y$-axis indicates the sensitivity value, reflecting the relative importance of each feature.
    }
    \label{fig:sensitivity}
\end{figure}

When constructing the dataset, we curated a selection of closely related features, such as heat capacity at constant volume ($C_{\rm v}$) and heat capacity at constant pressure ($C_{\rm p}$), as well as isothermal bulk modulus ($K_{\rm i}$) and static (adiabatic) bulk modulus ($K_{\rm s}$). The purpose of this approach was to ensure that the ML models achieve reliable accuracy. The correlation coefficient heatmap can reveal the potential degree of linear association between features, as illustrated in Fig.~\ref{fig:cor-map}. It is important to clarify that $K_{\rm i}$ and $K_{\rm s}$ are not entirely identical concepts~\cite{RN310,RN312,RN311}. Both are temperature-dependent, and the difference between them increases as the temperature rises, but due to the minimal difference at 300K, a near-linear relationship appears, which is also observed between the two specific heat capacities~\cite{RN310}. Interestingly, in addition to the previously mentioned bulk moduli and heat capacities, the Debye temperature $\Theta $, vibrational free energy per atom $F$, and vibrational entropy per atom ${S_{\rm{v}}}$ also show a high degree of linear correlation. The quantitative relationship between vibrational free energy and vibrational entropy is defined by the following equation~\cite{RN306}:
\begin{align}
    F = Q - T{S_{\rm{v}}},
    \label{eq:2}
\end{align}
where $Q$ represents the total enthalpy of formation of the compound,   denotes Kelvin temperature (here $T=300 {\rm K}$), and the equation above can be rewritten as $\frac{F}{{{S_{\rm{v}}}}} = \frac{Q}{{{S_{\rm{v}}}}} - T$. Due to the presence of the entropy-enthalpy compensation~\cite{RN331} effect, a linear relationship exists between entropy and enthalpy, which is further reflected in the linear correlation between $S_{\rm v}$ and $F$. Based on the work of G. D. Garbulsky and G. Ceder~\cite{RN309}, within the harmonic approximation, when the temperature is greater than the system's characteristic Debye temperature, the lattice Hamiltonian can be written as:
\begin{align}
    H\left( {\vec \sigma ,T} \right) = & {E_0}\left( {\vec \sigma } \right) + \left\langle {\ln \left( \omega  \right)} \right\rangle \left( \sigma  \right){k_{\rm{B}}}T \notag \\ 
    & + \frac{{\hbar \left\langle {{\omega ^2}} \right\rangle \left( {\vec \sigma } \right)}}{{24{k_{\rm{B}}}T}} 
    - \frac{{{\hbar ^4}\left\langle {{\omega ^4}} \right\rangle \left( {\vec \sigma } \right)}}{{2880k_{\rm{B}}^3{T^3}}} +  \cdots,
    \label{eq:3}
\end{align}
where the $\vec \sigma $ denotes the configuration of $A$ and $B$ atoms on the lattice, while $\left\langle  \cdot \right\rangle$ represents the average operation (per atom) over the Brillouin zone. ${E_0}$ refers to the fully relaxed ground state energy, $\omega $ is the vibrational frequency of a phonon mode, ${k_{\rm{B}}}$ and $\hbar $ correspond to the Boltzmann and Planck constants, respectively. As the temperature approaches the Debye temperature, setting $T = \Theta $, we have:
\begin{align}
    F \approx \left\langle {\ln \left( \omega  \right)} \right\rangle \left( \sigma  \right){k_{\rm{B}}}T,
    \label{eq:4}
\end{align}
where $\left\langle \ln \left( \omega  \right) \right\rangle \left( \sigma  \right)$ is a constant, thus the vibrational free energy per atom is approximately linearly related to the Debye temperature~\cite{RN309}. These conclusions, combined with Eq.~(\ref{eq:4}) sufficient explanation for the linear relationship between $\Theta $, $F$, and ${{S}_{\text{v}}}$.

We propose that among highly linearly correlated physical features, retaining only one is sufficient. Under ideal conditions, this approach can simplify the ML model without significantly degrading its performance. To achieve this goal, it is necessary to combine correlation analysis with sensitivity analysis. The purpose of sensitivity analysis is to determine which features have a greater impact on the target compared to others, or, more simply, which features are more ``important'' for the target. Currently, mainstream sensitivity analysis algorithms include SHAP~\cite{RN265}, LIME~\cite{RN269}, and Sobol~\cite{RN270}. However, these methods generally assume feature independence and the features used in our work are interdependent, their applicability is limited. Ignoring input interactions and multivariate distribution characteristics can severely skew or even invalidate any sensitivity analysis results~\cite{RN313}. Kucherenko et al.~\cite{RN314} improved the traditional Sobol index by using Copula-based sampling to separate each feature's marginal and joint distributions, thereby constructing a dependency model. In this work, we will implement the Kucherenko indices using UQLab~\cite{RN315} and compare it with other indices.

The assumption in Fig.~\ref{fig:sensitivity}(a) and (c) is that SHAP, LIME, and Sobol indices treat the input features as independent, which leads to qualitatively incorrect conclusions. This issue arises because Fig.~\ref{fig:cor-map} shows that the feature groups $\{F,S_{\rm v},\Theta\}$, $\{C_{\rm p},C_{\rm v}\}$, and $\{K_{\rm s},K_{\rm i}\}$ exhibit high mutual correlations. Considering that strongly correlated features are expected to exhibit similar sensitivities in the decision-making process~\cite{RN316}, the independence assumption becomes problematic. A more specific issue can be observed in Fig.~\ref{fig:sensitivity}(c), where, due to the dimensional constraint of the SISSO model (see Eq.~(\ref{eq:1})), only features $\{\Theta,\gamma,C_{\rm p},C_{\rm v},K_{\rm i},S_{\rm v}\}$ are retained and involved in the model construction. Even under this premise, both $S_i$ and $S_i^{\rm T}$ for $\{C_{\rm p},C_{\rm v}\}$ are exactly zero with respect to the remaining excluded variables. This is attributed to the presence of the term $\left| {{C_{\rm{p}}} - {C_{\rm{v}}}} \right|$ in Eq.~(\ref{eq:1}). From a global perspective, $C_{\rm p}$ and $C_{\rm v}$ exhibit nearly synchronous variations, resulting in an almost invariant value for their difference, which highlights a fundamental limitation of global sensitivity indices. In contrast, the local sensitivity metrics, such as SHAP and LIME, yield more reasonable interpretations. However, a remaining issue is that the sensitivity of $\gamma$ is evaluated to be zero—reflected in the LIME index where it shows sensitivity comparable to those of excluded variables. While the cause of this phenomenon remains unclear, it is evident that sensitivity analysis that neglects feature dependence yields qualitatively incorrect results for both KAN and SISSO-based surrogate models.
\begin{figure}[h]
    \centering
    \includegraphics[width=0.7\columnwidth]{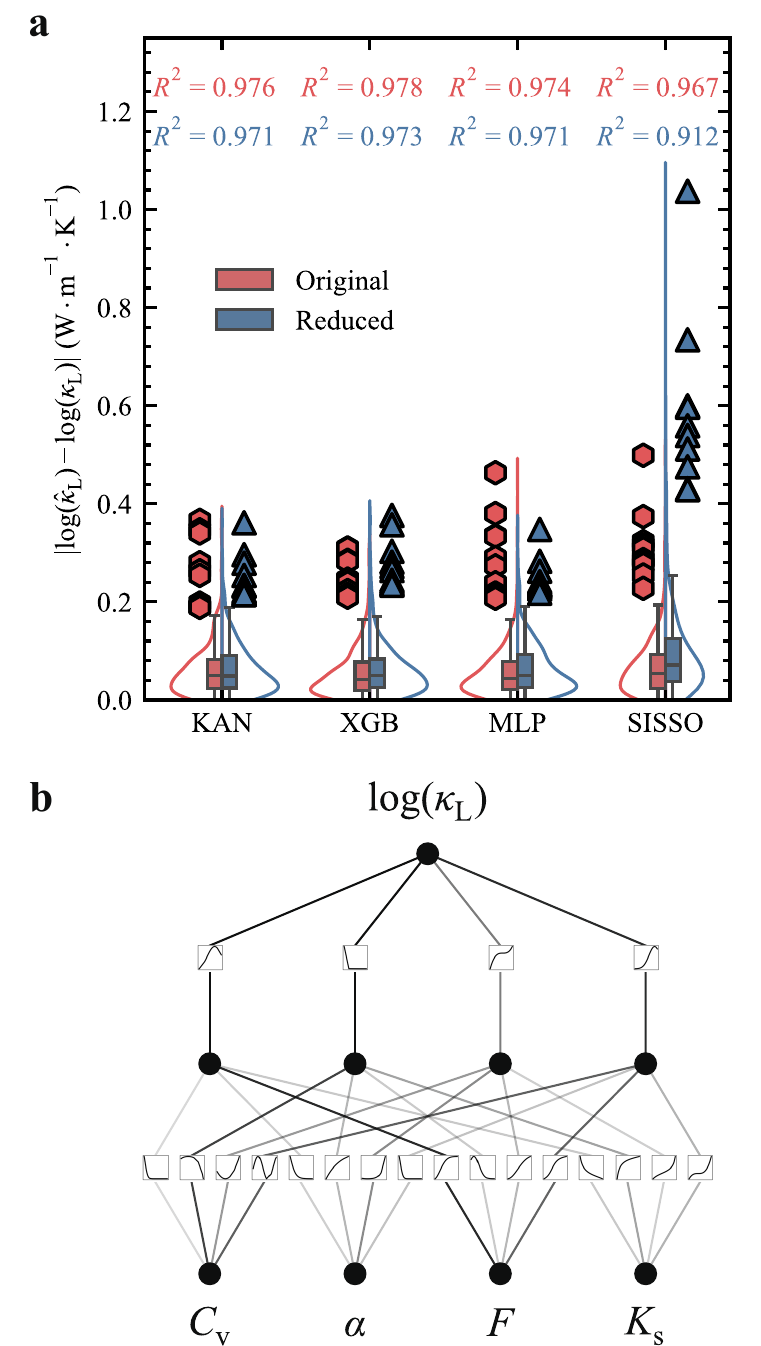}
    \caption{Performance comparison before and after dimensionality reduction. (a) After extracting key features and re-modeling the $\kappa_{\rm L}$, both KAN and the black-box models exhibited good robustness, while the robustness of SISSO did not meet expectations. (b) The simplified architecture of KAN model.}
    \label{fig:fig4}
\end{figure}

By contrast, the conclusions drawn from Kucherenko indices appear far more physically plausible, though the choice of surrogate model still introduces subtle differences. In Fig.~\ref{fig:sensitivity}(c), the individual contributions of features are quantified by the first-order index $K_{1}$. When KAN is used as the surrogate, all strongly correlated physical variables are assigned nearly identical sensitivity values, owing to the superior ability of Kucherenko indices to capture inter-feature dependencies. Moreover, the vanishing $K_{1}$ for ${\rm SG}$ aligns well with physical intuition, since existing models of $\kappa_{\rm L}$ do not explicitly incorporate space group symmetry~\cite{WANG2024101549,slack1979thermal,callaway1959model}. Interestingly, all features exhibit similar total sensitivity $K^{\rm T}$, which stems from the fully connected architecture of KAN: each feature contributes equally to the modelling of $\log(\kappa_{\rm L})$, rendering them structurally and interactively equivalent~\cite{RN264,RN156}.

Fig.~\ref{fig:sensitivity}(d) shows the outcome when SISSO is used as the surrogate. Compared to Fig.~\ref{fig:sensitivity}(c), the improvement lies in the ability of Kucherenko indices to assign similar sensitivity values to correlated but unmodelled inputs. However, due to the inherent limitations of the SISSO model, the contributions of $\{\gamma,C_{\rm v},C_{\rm p}\}$ remain underestimated. Taken together, only the combination of KAN and Kucherenko indices yields sensitivity results consistent with physical prior knowledge in our case.
\begin{table*}
    \centering
    \caption{The predicted results of $\kappa_{\rm L}$ based on KAN. The underlined materials have available DFT and MD calculation results.}
    \label{tab:3}
    \begin{tabular}{llccccc}
    \toprule
    \textbf{MPID} & \textbf{Formula} & \boldmath$F_{\rm pred}$ & \boldmath$K_{\rm s-pred}$ & \boldmath$\alpha_{\rm pred}$ & \boldmath$C_{\rm v-DFT}$ & \boldmath$\kappa_{\rm L-KAN}$ \\ 
    \midrule
    mp-1009818 & \underline{CN$_2$}     & 152.8788  & 333.6628  & $1.71 \times 10^{-5}$ & 0.2343  & 155.2907 \\
    mp-27710   & \underline{CrB$_4$}    & 113.7286  & 259.3992  & $2.73 \times 10^{-5}$ & 12.7181 & 29.3775  \\
    mp-1569    & \underline{Be$_2$C}    & 138.8222  & 191.1319  & $2.41 \times 10^{-5}$ & 4.6692  & 80.5562  \\
    mp-1096940 & CuBO$_2$   & 78.15269  & 267.6422  & $3.04 \times 10^{-5}$ & 8.4535  & 31.3254  \\
    mp-1183445 & BeSiO$_3$  & 85.25379  & 224.8873  & $2.56 \times 10^{-5}$ & 9.5514  & 41.3833  \\
    mp-1184997 & \underline{KRbAu$_2$}  & $-105.1144$    & 14.1935    & $2.36 \times 10^{-4}$ & 36.4862    & 0.2169        \\
    mp-1097263 & Cs$_2$RbNa & $-144.966$  & 16.8264   & $4.90 \times 10^{-4}$ & /       & /        \\
    mp-10378   & Cs$_3$Sb   & $-143.5$    & 11.2367   & $2.42 \times 10^{-4}$ & /       & /        \\
    mp-1097633 & Cs$_2$KRb  & $-145.791$  & 14.89677  & $5.10 \times 10^{-4}$ & /       & /        \\
    mp-635413  & Cs$_3$Bi   & $-168.671$  & 13.74844  & $2.50 \times 10^{-4}$ & /       & /        \\ 
    \botrule
    \end{tabular}
    \end{table*}
    
Ultimately, we retained $C_{\rm v}$, $\alpha$, $F$, and $K_{\rm s}$ as the features for the reduced dataset, as these features preserve as much of the original information as possible while enabling fast and accurate predictions via CGCNN, thereby facilitating high-throughput screening of new materials. As shown in Fig.~\ref{fig:fig4}(a), the four selected features are sufficient to describe the physical mechanism of $\kappa_{\rm L}$ with a high degree of confidence. Similar to other black-box models, KAN exhibits only a negligible decline in performance with the reduction in feature count. In contrast, SISSO's performance deteriorates significantly. This decline is likely due to the complexity constraints of the SISSO model. In contrast to KAN, SISSO is characterized by a relatively limited number of operators (see Eq.~(\ref{eq:eq13})), and its descriptors typically have a dimensionality of no more than 3. Exceeding this limit would lead to much greater computational resource demands compared to KAN and black-box models. These complexity constraints limit SISSO's expressive capacity and hinder its ability to capture complex physical feature-mapping relationships.
\begin{figure*}
    \centering
    \includegraphics[width=\textwidth]{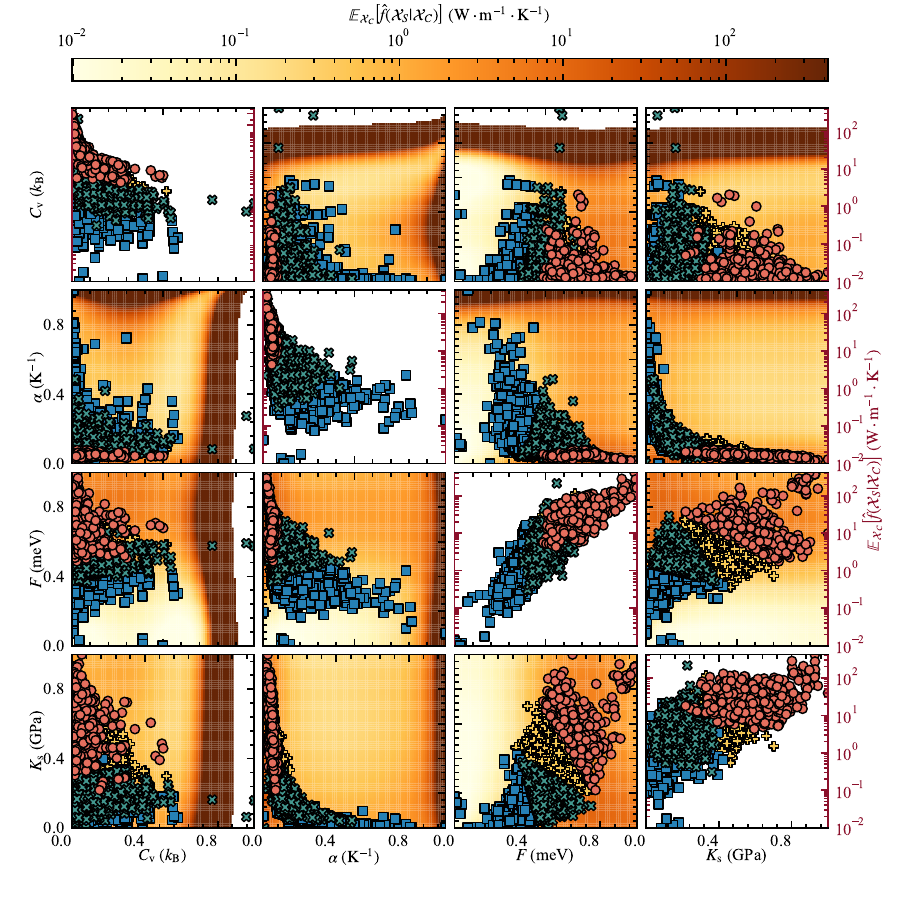}
    \caption{Partial dependency plots between each feature and $\kappa_{\rm L}$. In the diagonal subplots, the horizontal axis corresponds to the values of the feature shown at the bottom, while the red vertical axis represents the values of $\kappa_{\rm L}$. In the off-diagonal subplots, each scatter point's color and shape correspond to different clusters defined in $k$-means clustering, and the background color indicates the estimated thermal conductivity $\mathbb{E}_{\mathcal{X}_C}\left[\hat{f}(\mathcal{X}_S|\mathcal{X}_C)\right]$. Here, $\mathbb{E}_{\mathcal{X}_C}\left[\hat{f}(\mathcal{X}_S|\mathcal{X}_C)\right]$ refers to the predicted $\kappa_{\rm L}$ value when only the selected features $\mathcal{X}_S$ are varied, while all other features $\mathcal{X}_C$ are fixed at their dataset-wide mean values. For example, if $\mathcal{X}_S = \{F, K_{\rm s}\}$, then $\mathcal{X}_C = \{C_{\rm v}, \alpha\}$. The estimation $\mathbb{E}_{\mathcal{X}_C}\left[\hat{f}(\mathcal{X}_S|\mathcal{X}_C)\right]$ is intended to infer the $\kappa_{\rm L}$ in a virtual space where real samples are absent, based on the KAN-derived symbolic model.}
    \label{fig:fig5}
\end{figure*}

After key feature extraction, we performed symbolic regression again using KAN and SISSO. The symbolic model fitted by KAN became considerably simplified, as shown below:
\begin{align}
    \log {\left(\kappa_{\rm L}\right)_{{\rm{KAN}}}} = & - 1.05\sin d_1 + 0.61\cosh d_2 - 1.44 \notag \\ 
    & + 1.4{{\rm{e}}^{ - 19.72 d_3^2}} - 0.48{{\rm{e}}^{ - 63.21d_4^2}},
    \label{eq:5}
\end{align}
where
\begin{align*} 
    d_1 &= - 5.66{\left( {0.77 - F} \right)^2} - 0.7\tan \left( {1.56{K_{\rm s}} - 4.4} \right) + 2.89 \\
    &\quad + 1.56{e^{ - 13.65{{\left( { - \alpha  - 0.17} \right)}^2}}} + 0.85{{\rm{e}}^{ - 26.37{{\left( { - {C_{\rm v}} - 0.11} \right)}^2}}}, \\ 
    d_2 &= - 0.76\sin \left( {4.23F + 4.79} \right) + 1.03\tan \left( {1.55\alpha  - 0.99} \right)\\
    &\quad- 0.86\cosh \left( {5.35{C_{\rm v}} - 1.75} \right) + 2.31\\
    &\quad + 0.38{\mathop{\rm atan}\nolimits} \left( {8.56{K_{\rm s}} - 0.63} \right), \\ 
    d_3 &= {\left( {0.38 - {K_{\rm s}}} \right)^3} - 0.09\sin \left( {6.2{C_{\rm v}} - 6.16} \right)+ 0.47\\
    &\quad - 0.08{\mathop{\rm atan}\nolimits} \left( {4.8F - 1.66} \right) - 0.26{{\rm{e}}^{ - 38.44{{\left( { - \alpha  - 0.06} \right)}^2}}},\\ 
    d_4 &= - {\left( {0.37 - {C_{\rm v}}} \right)^2} - 0.13\tan \left( {1.46{K_{\rm s}} - 0.63} \right) - 0.3 \\
    &\quad + 0.28{e^{ - 2.53{{\left( { - F - 0.02} \right)}^2}}}.
\end{align*}

The analytical expression fitted by SISSO remains consistently concise:
\begin{align}
    \log {\left( \kappa_{\rm L}  \right)_{{\rm{SISSO}}}} = 1.53 + 0.0079\frac{{F\log \left( {{C_{\rm{v}}}} \right)}}{{\sqrt[3]{{{C_{\rm{v}}}}}}}  - 45.57\sqrt[3]{{{\alpha ^2}{C_{\rm{v}}}{K_{\rm{s}}}}}.
    \label{eq:6}
\end{align}
\begin{figure}[h]
    \centering
    \includegraphics[width=\columnwidth]{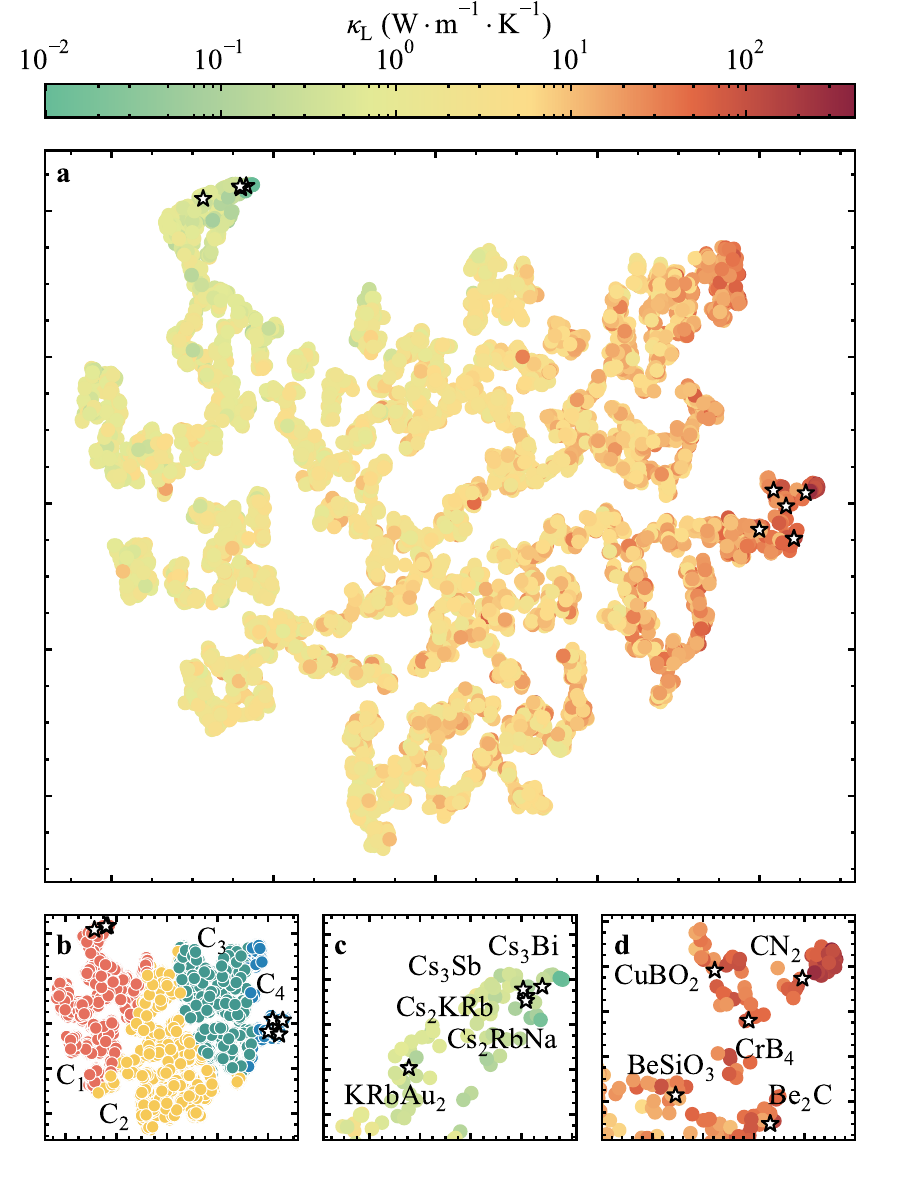}
    \caption{$k$-means clustering and $t$-SNE 2D visualization. (a) Labeled training samples from AFLOW are colored according to their $\kappa_{\rm L}$, with white pentagrams representing unlabeled candidates from the MP database. (b) The same scatter plot as in (a), but the samples are clustered into 4 groups, each represented by a different color and shape. (c) and (d) are zoomed-in views showing the thermal insulator/thermal conductor candidates in the $t$-SNE space. The $x$-axis and $y$-axis represent the two components of the $t$-SNE embedding, respectively.}
    \label{fig:tsne}
\end{figure}

Following feature reduction, the $R^2$ of the analytical expression derived by KAN was 0.9639, exhibiting minimal degradation in accuracy compared to its pre-reduction value of 0.9655. In contrast, as illustrated in Fig.~\ref{fig:fig4}(a), SISSO experienced a pronounced decline in accuracy. Consequently, we assert that KAN more effectively captures the mapping relationship between $\kappa_{\rm L}$ and the features, while also demonstrating exceptional robustness.
\begin{figure}[h]
    \centering
    \includegraphics[width=1\columnwidth]{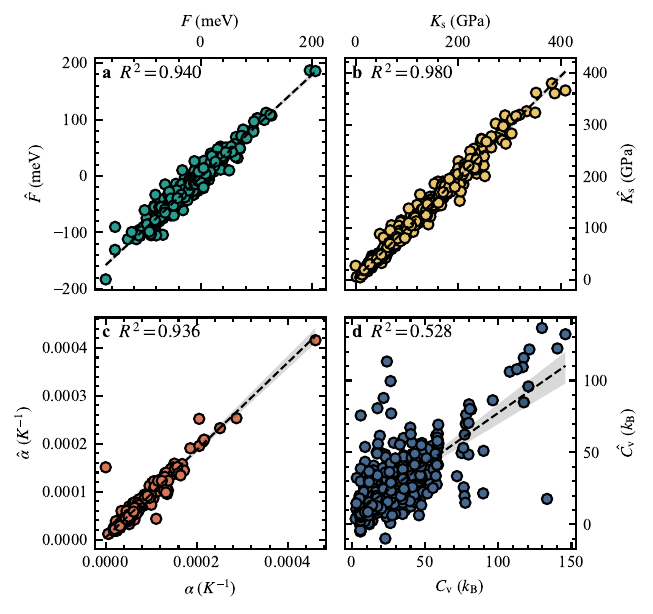}
    \caption{Predicting key features and qualitatively assessing $\kappa_{\rm L}$ via CGCNN. (a), (b), and (c) display parity plots comparing the CGCNN-predicted values of three features with the DFT-calculated values on their respective test sets, where $R^2_F=0.940$, $R^2_{K_{\rm s}}=0.980$, $R^2_{\alpha}=0.910$, and $R^2_{C_{\rm v}}=0.528$.}
    \label{fig:fig6}
\end{figure}

\subsection{Screening potential thermal insulators/conductors}

Unlike straightforward descriptors derived from composition (e.g., Magpie~\cite{RN317}) or structure (e.g., Coulomb matrix~\cite{RN318}), the physical features in the reduced dataset are more intricate and not readily computable from fundamental physical quantities. Inspired by previous studies~\cite{RN320,RN319}, our approach is to use a two-stage prediction method. We predict primary physical features by CGCNNs, and then use the predicted primary features to perform high-throughput predictions of the $\kappa_{\rm L}$ for unlabeled materials. CGCNNs have achieved accuracy comparable to or even surpassing DFT in predicting energy-related and mechanical properties~\cite{RN321}. In our application, CGCNNs delivered fairly accurate results for the prediction of all primary features except for $C_{\rm v}$, where the prediction accuracy was notably lower (see Supplementary Information for details).
Therefore, we focus on first screening a subset of potential thermal insulating/conductive materials using primary features that can be reliably predicted qualitatively, and then calculating $C_{\rm v}$ using DFT. This approach is more efficient than directly using DFT or MD methods to calculate $\kappa_{\rm L}$.

Based on the information from Fig.~\ref{fig:cor-map} and Fig.~\ref{fig:fig5}, it can be concluded that $\kappa_{\rm L}$ is strongly and positively correlated with both $F$ and $K_{\rm s}$. Materials with higher $F$ and $K_{\rm s}$ values generally exhibit higher $\kappa_{\rm L}$, a trend further confirmed by the $F$–$K_{\rm s}$ scatter plot in Fig.~\ref{fig:fig5}. Therefore, the preliminary qualitative screening of thermal insulators and conductors based on $F$ and $K_{\rm s}$ is highly beneficial for accelerating the exploration of new materials. The predictions of $F$, $K_{\rm s}$ and $\alpha$ using CGCNN are shown in Fig.~\ref{fig:fig6}(a)-(c), with $R^2$ values reaching 0.94, 0.98 and 0.91, respectively. We extracted a total of 2,246 samples from the Materials Project (MP)~\cite{RN322} with the following selection criteria:
\begin{itemize}
    \item Excluding transition metal elements;
    \item Number of atoms (n$_{\rm sites}$) $<$ 5;
    \item Band gap $ \in \left( {0,1.5{\rm{eV}}} \right]$.
\end{itemize}

Although CGCNN is capable of accurately evaluating $F$, $K_{\rm s}$, and $\alpha$ for unlabeled materials, to the best of our knowledge, there is currently no reliable machine learning model that exhibits outstanding performance in predicting $C_{\rm v}$. When performing a quantitative estimation of $\kappa_{\rm L}$ for a large number of new materials based on Eq.~(\ref{eq:5}), the $C_{\rm v}$ parameter still needs to be obtained from DFT calculations. However, this approach imposes a considerable computational cost when dealing with a large number of samples. To address this, we carry out unsupervised learning to rapidly screen the most promising thermal conductors and insulators from the MP database, enabling precise identification of candidate materials.

For this purpose, we first attempt to use $k$-means clustering~\cite{sinaga2020unsupervised} on samples from the AFLOW database (e.g., the labeled dataset used for model training in Section~\ref{model-training}) to evaluate its effectiveness. Since reliable quantitative estimation is currently only feasible for the three primary features $F$, $K_{\rm s}$, and $\alpha$, we retain only these three features for each sample during clustering. In $k$-means clustering, the optimal number of clusters is typically determined using the elbow method~\cite{cui2020introduction} or the silhouette score~\cite{shahapure2020cluster}. However, in our experiments, these two methods yielded inconsistent results, making it unclear which one is more convincing (see Supplementary Information). 
Fortunately, through manual inspection, we found that when $K=4$, the clustering results already exhibit a clear and meaningful trend.
\begin{figure}[h]
    \centering
    \includegraphics[width=\columnwidth]{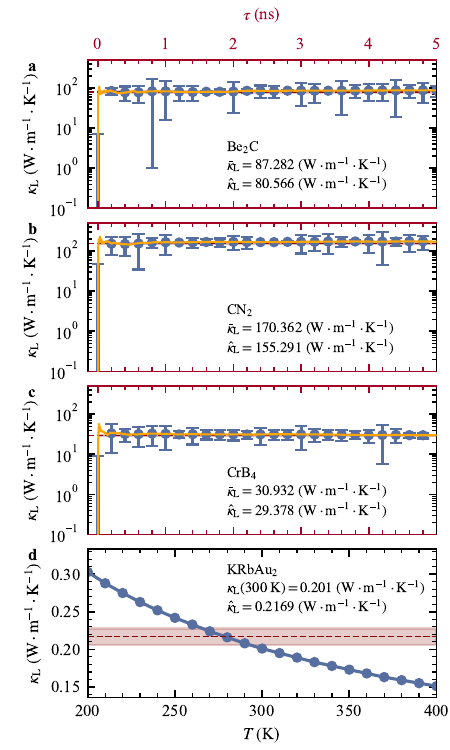}
    \caption{MD and DFT validation of candidates' $\kappa_{\rm L}$. For (a) Be$_2$C, (b) CN$_2$, and (c) CrB$_4$, the MD results are shown as blue points with error bars, together with the running average curve (orange). The deep red dashed line denotes the $\hat{\kappa}_{\rm L}$ predicted by the KAN model, and the shaded red band indicates the $\pm5\%$ interval. The MD results demonstrate that a relaxation time of 5~ns is sufficient for convergence of $\kappa_{\rm L}$. (d) For KRbAu$_2$, the blue circles connected by lines show the temperature dependence of $\kappa_{\rm L}$ obtained from DFT, while the deep red dashed line and shaded band indicate the KAN prediction and its $\pm5\%$ error interval.}
    \label{fig:fig7}
\end{figure}

In Fig.~\ref{fig:tsne}(a), all training samples are distinctly clustered into seven clusters, $C_1$ to $C_4$, with well-defined boundaries between clusters. Mapping the values of $\log(\kappa_{\rm L})$ onto the classification results in Fig.~\ref{fig:tsne}(a), as shown in Fig.~\ref{fig:tsne}(b), we observe that as $t$-SNE component 1 increases, $\log(\kappa_{\rm L})$ exhibits an upward trend. This allows us to identify $C_1$ as the cluster corresponding to thermal insulators and $C_4$ as the cluster representing thermal conductors. More importantly, the clustering results indicate that $\log(\kappa_{\rm L})$ can be accurately and qualitatively assessed using only the three features $F$, $K_{\rm s}$, and $\alpha$. Based on the labeled dataset, we obtained a CGCNN pre-trained model for the initial features and applied it to predict $F$, $K_{\rm s}$, and $\alpha$ for unlabeled samples from the MP database. These predicted features were then clustered in the same manner as the labeled samples. 
In Fig.~\ref{fig:tsne}(b), we selected the 10 most promising thermal insulator/conductor candidates from clusters $C_1$ and $C_4$ (see Table~\ref{tab:3}). The zoomed-in views are shown in Figures~\ref{fig:tsne}(c) and (d), where these candidates are found in the same clusters as the labeled samples with the lowest/highest $\log(\kappa_{\rm L})$, respectively. Experimental results proved that only 4 out of the 10 candidates (CN$_2$, CrB$_4$, Be$_2$C and KRbAu$_2$) effectively utilize DFT and MD to calculate $C_{\rm v}$ and $\kappa_{\rm L}$, respectively. The subsequent validation process will also focus on these 4 feasible structures.

Among the thermal conductor candidates, only Be$_2$C belongs to the cubic crystal system. CN$_2$ exhibits high thermal conductivity along the $z$-direction while being nearly insulating in the $x$- and $y$-directions. CrB$_4$ demonstrates anisotropic thermal conductivity in all three directions. Following the approach in relevant studies~\cite{pereira2016anisotropic,termentzidis2018thermal}, we employed MD simulations to validate the $\kappa_{\rm L}$ values of these conductor candidates. For Be$_2$C, calculations are required for only one direction due to its cubic symmetry. For CN$_2$, we considered only $\kappa_{\rm L}$ along the $z$-axis. In the case of CrB$_4$, the overall $\kappa_{\rm L}$ is determined as the average of the values computed along the three principal directions. 

According to our MD calculations, the actual $\kappa_{\rm L}$ values of the candidate materials fall within distinct ranges ($\sim 30$, $\sim 80$, $\sim 150$ $\rm{W \cdot m^{-1} \cdot K^{-1}}$). As shown in Fig.~\ref{fig:fig7}(e), samples with $\kappa_{\rm L} > 100$ $\rm{W \cdot m^{-1} \cdot K^{-1}}$ account for only 0.47\% of the entire training dataset. In such cases of data imbalance, model predictions for regions with scarce samples typically exhibit considerable uncertainty~\cite{khan2019striking,sun2024data}.
\begin{figure*}
    \centering
    \includegraphics[width=0.9\textwidth]{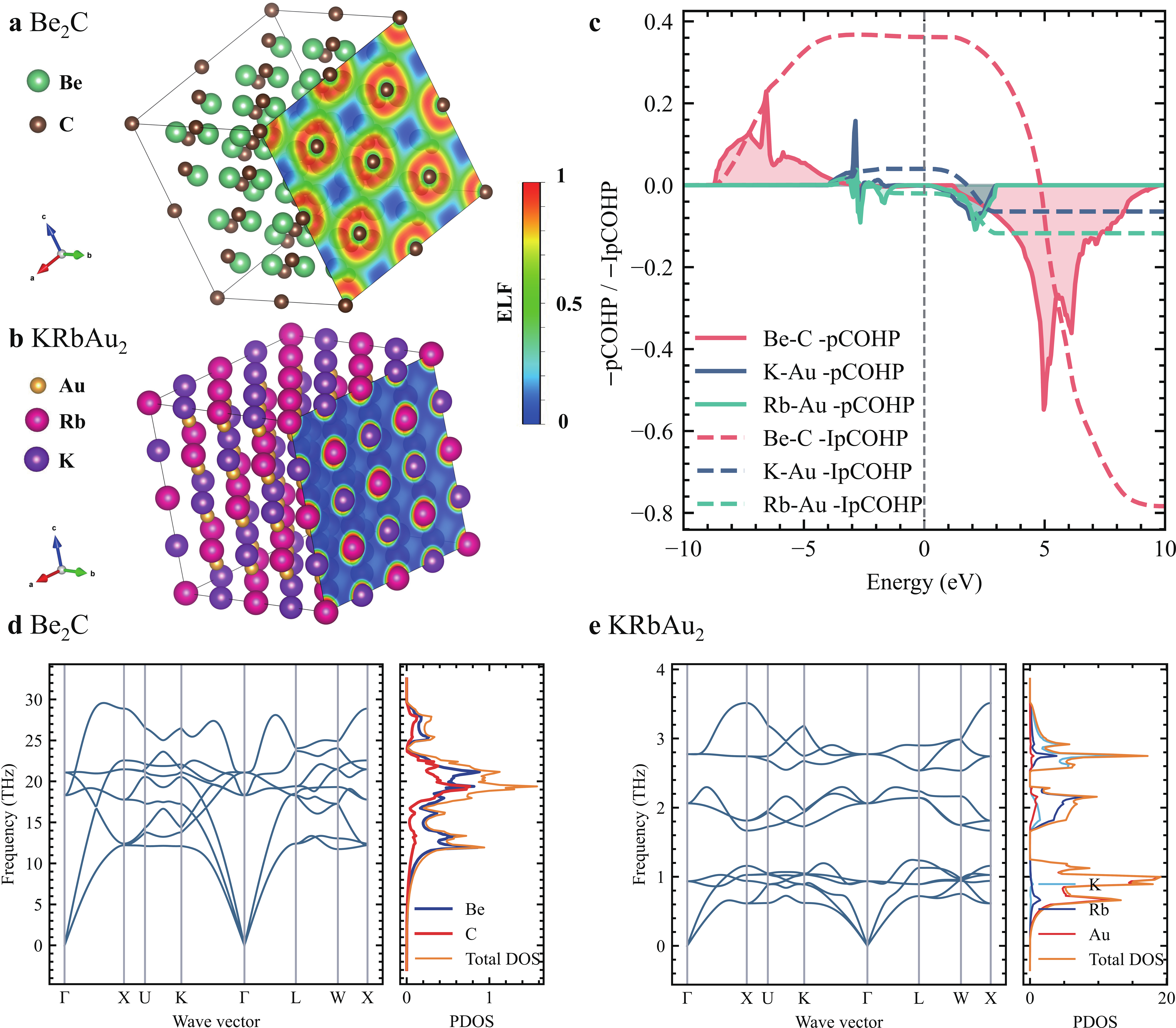}
    \caption{Based on DFT-based phonon thermal transport analysis. Crystal structures and the projected 2D electron localization function (ELF) diagram of (a) Be$_2$C and (b) KRbAu$_2$. (c) is the $-$pCOHP \& $-$IpCOHP for Be-C, K-Au and Rb-Au bonds. In (d) and (e), the left panels depict the phonon dispersion curves of Be$_2$C and KRbAu$_2$, while the right panels present the corresponding atom-projected PDOS.
    }
    \label{fig:pdos}
\end{figure*}

KRbAu$_2$ has the space group symbol $\textit{Fm}\overline{3}\textit{m}$, belonging to the cubic crystal system. For such isotropic thermal insulator candidates, it is generally more efficient and accurate to calculate $\kappa_{\rm L}$ using DFT~\cite{chen2019understanding,malakkal2017thermal}. As shown in Fig.~\ref{fig:fig7}(d), the DFT-calculated $\kappa_{\rm L}$ for this structure at 300K is 0.201 $\rm{W \cdot m^{-1} \cdot K^{-1}}$, while the KAN-predicted value is 0.2169, correctly identifying it as a thermal insulator. In summary, the methods of $k$-means clustering and $t$-SNE visualization are also sufficient for qualitatively assessing the $\kappa_{\rm L}$ of materials, enabling high-throughput screening in combination with the pretrained model. Furthermore, the $\kappa_{\rm L}$ values predicted by KAN exhibit excellent agreement with both MD and DFT validation results. This demonstrates that the pretrained KAN model achieves not only reliable accuracy but also satisfactory extrapolation capability.

\subsection{Tracing high/low LTC from the perspective of theoretical calculation}

Among the validated candidates, considering the isotropy of Be$_2$C and KRbAu$_2$, we take these two as examples to explore the mechanisms influencing high and low $\kappa_{\rm L}$ based on DFT. Their crystal structures are shown in Figures~\ref{fig:pdos}(a) and (b), both belonging to the cubic system but exhibiting distinctly different phonon transport behaviors. Be$_2$C adopts a [BeC$_4$] tetrahedral configuration, where each C atom is coordinated with four Be atoms, with a bond length of approximately 1.64 ${\rm \AA}$. These [BeC$_4$] tetrahedral units share edges, forming a continuous three-dimensional covalent network that imparts high rigidity and thermal stability to the crystal. In contrast, KRbAu$_2$ consists of K, Rb, and Au, forming a metal-ion mixed-bond network. The Au atoms exist as [Au$_2$] dimers, where each Au atom is connected via metallic bonds to form a framework structure. K and Rb act as cations filling the interstitial spaces, providing electrostatic stabilization. Due to the large atomic radii of K and Rb, the introduced steric hindrance effect increases lattice flexibility~\cite{guo2022atomistic}.
\begin{figure*}
    \centering
    \includegraphics[width=\textwidth]{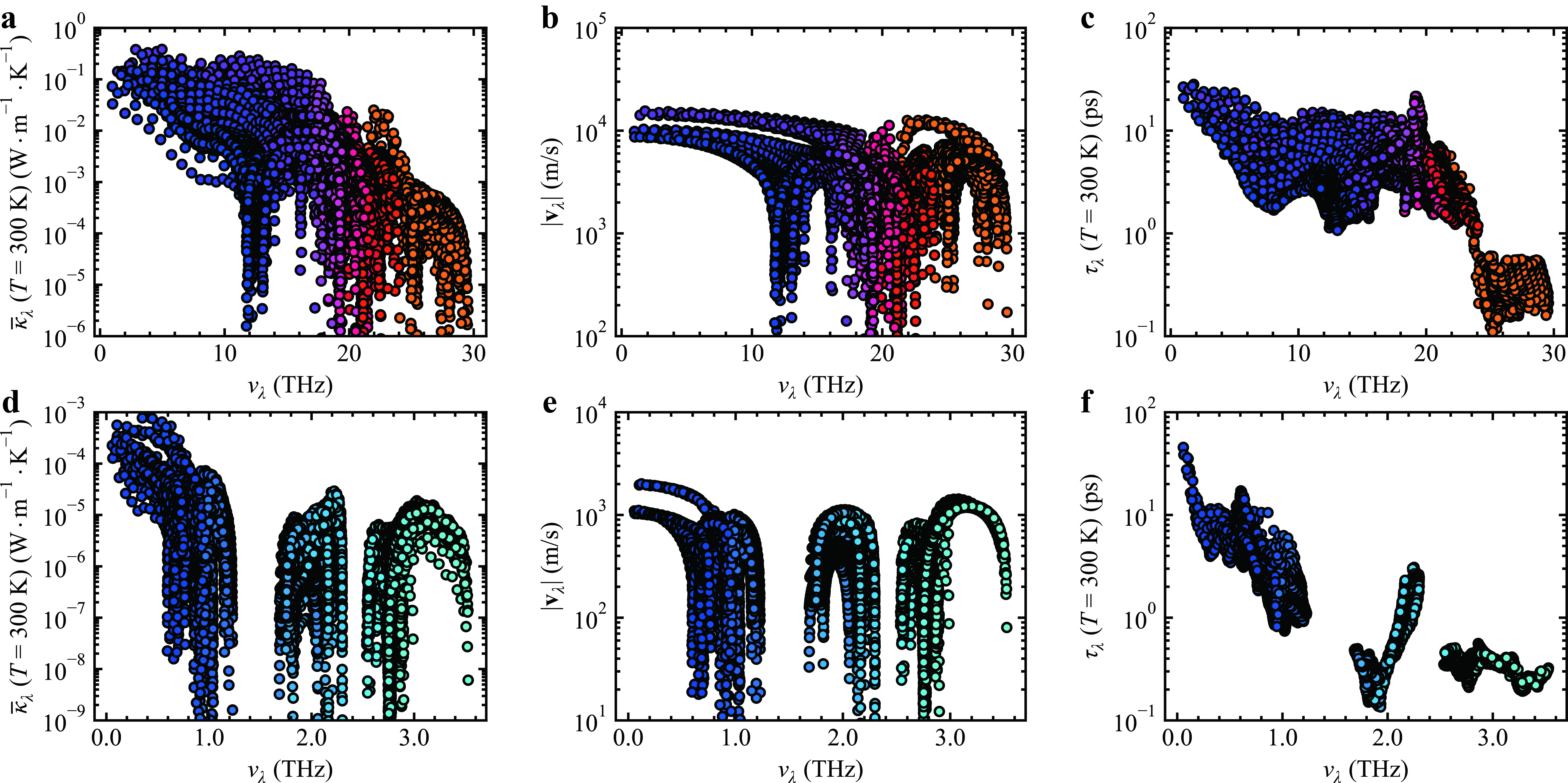}
    \caption{Modal quantities of Be$_2$C and KRbAu$_2$. (a-c) illustrate the frequency dependence of the average thermal conductivity $\bar \kappa_{\lambda}$, group velocity $|\mathbf{v}_{\lambda}|$, and phonon lifetime $\tau_{\lambda}$ for Be$_2$C, respectively; (d-f) present the counterparts for KRbAu$_2$. The scatters are colored according to the band index, with two distinct colormaps used to differentiate the two materials.}
    \label{fig:kappa-gv-tau}
\end{figure*}

The effects of lattice rigidity and flexibility on $\kappa_{\rm L}$ primarily manifest in their influence on phonon group velocity $v_g$ and anharmonicity~\cite{PhysRevMaterials.5.053801}. In Be$_2$C, the strong covalent bonding between Be and C forms a highly rigid lattice, leading to smaller atomic displacements and lower lattice vibrational anharmonicity, thereby reducing three-phonon scattering (primarily suppressing Umklapp scattering)~\cite{maznev2014demystifying}. Additionally, such crystals typically exhibit high elastic moduli, resulting in steep phonon dispersion curves, which lead to higher phonon velocities and enhanced thermal conductivity~\cite{xie1999high}. In contrast, the softer bonds between K, Rb, and Au make the lattice more deformable. A more flexible lattice experiences larger thermal vibrations, inducing stronger three-phonon and four-phonon scattering, which significantly reduces $\kappa_{\rm L}$. Atomic mass also indirectly influences $\kappa_{\rm L}$ by affecting sound velocity. Be$_2$C consists of light atoms, whereas KRbAu$_2$ is composed entirely of heavy atoms. Specifically, heavy atoms tend to reduce group velocity, thereby exhibiting thermal insulation characteristics, whereas light atoms have the opposite effect~\cite{amzongshu,aszongshu,jain2014thermal}.

To further investigate the impact of chemical bonding on the $\kappa_{\rm L}$ of Be$_2$C and KRbAu$_2$, we employ the Negative Projected Crystal Orbital Hamilton Population ($-$pCOHP) to visualize their local bonding characteristics. In $-$pCOHP, positive values indicate bonding states, negative values represent antibonding states, and a value of zero signifies non-bonding interactions. 
As shown in Fig.~\ref{fig:pdos}(c), the Fermi level $E_{\rm F}$ lies at 0 eV. For the Be-C bond, bonding states dominate below $E_{\rm F}$ down to -8.7 eV. Strong bonding implies greater lattice rigidity and a steeper potential energy surface~\cite{schlegel2003exploring}, leading to higher optical branch vibration frequencies, which suppress phonon scattering and enhance thermal conductivity. This behavior is similar to classical high-$\kappa_{\rm L}$ materials such as diamond and SiC. In KRbAu$_2$, the bonding characteristics differ between the two types of bonds: K-Au is bonding, while Rb-Au is antibonding. The antibonding nature of the Rb-Au bond softens the lattice, enhances phonon scattering, and reduces $\kappa_{\rm L}$. Fig.~\ref{fig:fig7}(d) confirms the ultra-low $\kappa_{\rm L}$ of KRbAu$_2$, suggesting that its phonon transport properties are primarily dictated by the Rb-Au bond. 
The $-$IpCOHP analysis further evaluates global bond strength. Despite K-Au being a bonding interaction, its bond strength remains weak, and the presence of an antibonding Rb-Au bond exacerbates phonon anharmonicity. Weak chemical bonds are often associated with strong phonon anharmonicity~\cite{lee2014resonant,li2015orbitally}, which is a key factor underlying the ultra-low $\kappa_{\rm L}$ of KRbAu$_2$.

Figures~\ref{fig:pdos}(d) and (e) provide insights into the characteristics of low-energy optical branches and acoustic modes, supporting the previous hypotheses. 
In the phonon dispersion spectrum of Be$_2$C, all phonon modes exhibit real frequencies, indicating dynamical stability at zero temperature with no tendency for structural collapse. The acoustic branches in the low-frequency region ($0 \sim 10$ THz) show relatively high group velocities, suggesting potentially high phonon thermal conductivity. In contrast, the optical branches are primarily distributed in the high-frequency range (15 $\sim$ 30 THz) and exhibit a certain degree of flatness, indicating lower group velocities and minimal contribution to heat transport. The Projected Density of State (PDOS) analysis on the right side of Fig.~\ref{fig:pdos}(d) further reveals that heat-carrying phonons are predominantly below 15 THz, mainly contributed by the heavier Be atoms, while the high-frequency optical branches involve both C and Be atoms.
For KRbAu$_2$, the phonon dispersion curve shows that all modes remain within the real frequency range, with no imaginary branches, confirming its dynamical stability at zero temperature. The acoustic branches are concentrated in the low-frequency region (0 $\sim$ 1 THz) and exhibit flat dispersions, resulting in low group velocities. The optical branches are distributed between 1 and 3.5 THz, with some modes displaying significant dispersion. According to the PDOS, the dominant vibrating atoms shift sequentially from Au to Rb to K as frequency increases, consistent with the mass distribution of these elements.

Fig.~\ref{fig:kappa-gv-tau}(b) and (e) show the frequency dependence of the group velocity $\mathbf{v}_g$ at the center of the Brillouin zone for Be$_2$C and KRbAu$_2$, respectively. The group velocity of Be$_2$C exceeds 10 km/s, while that of KRbAu$_2$ does not exceed 1.1 km/s. This significant difference in group velocity can be attributed to differences in bonding and strength, which is consistent with the inference made from the steepness of their phonon spectra. Fig.~\ref{fig:kappa-gv-tau}(c) and (f) describe the differences in phonon lifetimes between the two materials. In the low-frequency range below 10 THz, Be$_2$C exhibits longer phonon lifetimes, with the longest exceeding 10 ps and the shortest being above 1 ps. In contrast, in KRbAu$_2$, the phonon lifetime decreases sharply with increasing frequency, with $\tau_{\lambda}$ approaching 0.1 ps at $\upsilon_{\lambda} = 2 {\rm THz}$. It is noteworthy that for certain specific low-frequency modes, the phonon lifetime $\tau_{\lambda}$ of KRbAu$_2$ exceeds that of Be$_2$C, which may be due to special scattering suppression effects, e.g., weaker three-phonon scattering~\cite{PhysRevX.10.021063}. However, overall, Be$_2$C has a longer and more uniform phonon lifetime, which is favorable for long-range heat transport, while the shorter phonon lifetime of KRbAu$_2$ suppresses $\kappa_{\rm L}$.

\section{Conclusion}

In the past, researchers often relied on empirical or semi-empirical models to calculate the $\kappa_{\rm L}$ of materials. However, as the range of explored materials continues to expand, the inaccuracies inherent in empirical models have become increasingly problematic. Machine learning-based $\kappa_{\rm L}$ modeling has achieved remarkable accuracy, but improvements in accuracy alone offers limited contributions to the advancement of materials science. Black-box models, such as neural networks and ensemble learning, excel in accuracy but their complex structures hinder the understanding of feature-target relationships. In contrast, white-box models like symbolic regression provide transparent structures, making their internal mechanisms interpretable, but this often comes at the cost of accuracy. Traditionally, the interpretability and accuracy of ML or DL models have been seen as mutually exclusive, akin to ``having your cake and eating it too.'' However, in the context of $\kappa_{\rm L}$ modeling, interpretable DL models like KAN have demonstrated significant success in balancing both accuracy and interpretability.

Our work employed KAN to model $\log(\kappa_{\rm L})$, demonstrating that its performance in terms of both accuracy and robustness is fully comparable to that of black-box models. For the interpretability analysis of white-box models, we considered feature interaction effects and dependencies. Sensitivity analysis results based on the KAN and SISSO models revealed that KAN can accurately extract key features of $\kappa_{\rm L}$, highlighting its superior interpretability in this context. We combined KAN with CGCNN to construct a two-stage prediction framework, where CGCNN predicts key features, and the KAN pretraining model maps these features to $\log(\kappa_{\rm L})$; for physical properties like $C_{\rm v}$, which CGCNN struggles to predict accurately, we used DFT calculations to complement the framework. In the case of unlabeled new samples, we applied $k$-means clustering to identify 10 potential thermal conductors/insulators. Among them, the $\kappa_{\rm L}$ values of CN$_2$, CrB$_4$, Be$_2$C, and KRbAu$_2$ were verified by DFT and MD, yielding results highly consistent with deep learning predictions, which sufficiently demonstrate KAN's excellent extrapolation capability. We visualized the phonon spectra, PDOS, $-$pCOHP, and other information of Be$_2$C and KRbAu$_2$ using DFT, analyzing the phonon transport mechanisms of both materials from the perspectives of chemical bonding, atomic mass, and structural stability, providing theoretical insights as reference.

\section{Future perspectives}

This work demonstrates the capability of KAN in predicting $\kappa_{\rm L}$, but several key issues remain to be explored for further advancing the research.

\textbf{Towards an All-Feature Predictor.} Among the four parameters describing $\kappa_{\rm L}$, although CGCNN can accurately estimate the vibrational free energy $F$, the static bulk modulus $K_{\rm s}$, and the thermal expansion coefficient $\alpha$ from the crystal structure file, to the best of our knowledge, there is currently no straightforward method to accurately determine the constant-volume heat capacity $C_{\rm v}$ from the crystal structure. Future research should explore more powerful network architectures or models incorporating physical priors to enhance the fitting capability for thermodynamic properties that are difficult to predict, thereby achieving a truly DFT-free, fully machine learning-based high-throughput screening workflow.

\textbf{Extension to Multi-Temperature Modeling.} Due to the limitation of available data, this work focuses on modeling and predicting $\kappa_{\rm L}$ at 300K. In reality, $\kappa_{\rm L}$ is a temperature-dependent variable. Future studies may consider building a comprehensive dataset encompassing temperature variation through DFT or MD calculations or experimental data collection to extend the current framework.

\textbf{Inverse Design.} The framework proposed in this work can be further applied to inverse design tasks: by optimizing the crystal structure, candidate materials with target $\kappa_{\rm L}$ values can be identified. Combined with generative models or evolutionary algorithms, this approach holds promise for the targeted discovery of high or low thermal conductivity materials.

\section*{Acknowledgement}
We would like to acknowledge the financial support from the Translational Medicine and Interdisciplinary Research Joint Fund of Zhongnan Hospital of Wuhan University (Grant No.~ZNJC202235, No.~ZNJC202424), the Hubei Provincial Key Technology Foundation of China (No.~2021ACA013), and the National Natural Science Foundation of China (No.~22327901). The numerical calculations in this work have been done on the supercomputing system in the Supercomputing Center of Wuhan University.

\section*{Data Availability}
The original LTC data and the selected features can be downloaded from \url{https://aflowlib.org/} or accessed via the GitHub repository at \url{https://github.com/FlorianTseng/LTC-modeling}.

\section*{Code Availability}
The codes supporting our research are available at \url{https://github.com/FlorianTseng/LTC-modeling}.

\section*{Author Contributions}

Yuxuan Zeng: Investigation, Methodology, Writing–original draft, Formal analysis, Data curation, Conceptualization, Visualization. Wei Cao: Writing-original draft, Methodology, Supervision, Resources, Funding acquisition, Project administration. Yijing Zuo: Data curation, Formal analysis, Methodology. Tan Peng: Data curation. Yue Hou: Project administration, Funding acquisition. Ling Miao: Software. Ziyu Wang: Supervision, Resources, Project administration, Funding acquisition, Writing-review \& editing. Jing Shi: Resources, Project administration, Funding acquisition.

\section*{Competing Interests}

The authors declare no competing interests.

\clearpage

\bibliography{ref}
\bibliographystyle{naturemag}

\end{document}